\documentclass[a4paper,english,nobibnotes,nofootinbib,notitlepage,superscriptaddress]{revtex4-1}
\usepackage[latin9]{inputenc}
\usepackage{amsmath}
\usepackage{amssymb}
\usepackage{graphicx}
\usepackage{tikz}
\usetikzlibrary{trees}
\usetikzlibrary{decorations.pathmorphing}
\usetikzlibrary{decorations.markings}
\usetikzlibrary{shapes.misc}

\makeatletter

\pdfpageheight\paperheight
\pdfpagewidth\paperwidth

\usepackage{amssymb,amsmath}
\usepackage{epsfig,subfigure}

\def\b#1{\mathbf{#1}}
\def\d{\partial}
\def\D#1{$\displaystyle{#1}$}
\def\f{\frac}
\def\h#1{\widehat{#1}}

\def\lp#1{\left(#1\right)}

\def\vev#1{\left\langle#1\right\rangle}
\def\w#1{\overline{#1}}
\def\FC{\mathcal{F}}
\def\HC{\mathcal{H}}

\newcommand{\be}{\begin{equation}}
\newcommand{\ee}{\end{equation}}
\newcommand{\bea}{\begin{eqnarray}}
\newcommand{\eea}{\end{eqnarray}}

\newcommand{\bra}{\langle}
\newcommand{\ket}{\rangle}
\newcommand\lr[1]{{\left({#1}\right)}}

\makeatother

\usepackage{babel}
\begin{document}

\title{Linearly-polarized small-x gluons in forward heavy-quark pair production}

\author{C. Marquet}\email{cyrille.marquet@polytechnique.edu}
\affiliation{Centre de Physique Th\'eorique, \'Ecole Polytechnique, CNRS, Universit\'e Paris-Saclay, F-91128 Palaiseau, France}

\author{C. Roiesnel}\email{claude.roiesnel@polytechnique.edu}
\affiliation{Centre de Physique Th\'eorique, \'Ecole Polytechnique, CNRS, Universit\'e Paris-Saclay, F-91128 Palaiseau, France}

\author{P. Taels}\email{pieter.taels@uantwerpen.be}
\affiliation{Centre de Physique Th\'eorique, \'Ecole Polytechnique, CNRS, Universit\'e Paris-Saclay, F-91128 Palaiseau, France}
\affiliation{Department of Physics, University of Antwerp, Groenenborgerlaan 171, 2020 Antwerpen, Belgium}
\affiliation{The H. Niewodnicza\'nski Institute of Nuclear Physics PAN, Radzikowskiego 152, 31-342 Krak\'ow, Poland}

\begin{abstract}

We use the Color Glass Condensate (CGC) framework to study the production of forward heavy quark-antiquark pairs in unpolarized proton-nucleus or proton-proton collisions in the small-$x$ regime. In the limit of nearly back-to-back jets, the CGC result simplifies into the transverse-momentum dependent (TMD) factorization approach. For massless quarks, the TMD factorization formula involves three unpolarized gluon TMDs: the Weizs\"{a}cker-Williams gluon distribution, the adjoint-dipole gluon distribution, and an additional one. When quark masses are kept non-zero, three new gluon TMDs appear -- each partnered to one of the aforementioned distributions -- which describe the distribution of linearly-polarized gluons in the unpolarized small-$x$ target. We show how these six gluon TMDs emerge from the CGC formulation and we determine their expressions in terms of Wilson line correlators. We calculate them analytically in the McLerran-Venugopalan model, and further evolve them towards smaller values of $x$ using a numerical implementation of JIMWLK evolution.

\end{abstract}

\maketitle

\section{Introduction}

In hadronic reactions that are governed by more than one hard momentum
scale, the standard QCD framework of collinear factorization at leading
twist becomes insufficient, and one needs to resort to more sophisticated
factorization schemes. One such scheme is TMD factorization \cite{Boer:1999si,Mulders:2000sh,Belitsky:2002sm,Boer:2003cm,Bomhof:2006dp,Vogelsang:2007jk,Collins:2007nk,Rogers:2010dm}, which makes use of transverse-momentum-dependent
parton distributions, or TMDs for short. One of the many intricacies
of TMDs is the fact that, in contrast to the usual collinear PDFs,
their operator definition depends on the hard process under consideration,
hence at first glance, universality is broken.

In recent years, many efforts have been made to elucidate the properties of TMDs in the high-energy or small-$x$ limit \cite{Marquet:2009ca,Dominguez:2010xd,Metz:2011wb,Dominguez:2011wm,Dominguez:2011gc,dominguez:2011br,Akcakaya:2012si,Kotko:2015ura,Petreska:2015rbk,Balitsky:2015qba,Balitsky:2016dgz,Zhou:2016tfe,vanHameren:2016ftb,Marquet:2016cgx}.
A process particularly adapted to this study is forward quark-antiquark pair production in high-energy proton-nucleus collisions. For kinematical reasons,
in such a process, the proton side of the collision involves large-$x$ partons, while on the nucleus side, small-$x$ gluons participate. Hence, this process can be described in a hybrid approach \cite{Dumitru:2005gt,Altinoluk:2011qy,Chirilli:2011km}, in which the proton content is described by regular, integrated
PDFs, while the small-$x$ dynamics in the nuclear wave function is dealt with using the Color Glass Condensate (CGC) effective theory
\cite{Balitsky:1995ub,JalilianMarian:1997jx,JalilianMarian:1997dw,Kovchegov:1999yj,Weigert:2000gi,Iancu:2000hn,Ferreiro:2001qy,Gelis:2010nm}.

More specifically, forward quark-antiquark pair production in dilute-dense collisions is characterized by three momentum scales: $P_{t}$, the typical transverse
momentum of a single quark, and always one of the largest scales; $k_{t}$, the total transverse momentum of the pair, which is a measure of the transverse momentum of the small-$x$ gluons coming from the target; and $Q_s$, the saturation scale of the nucleus, which is always one of the softest scales.
The value of $k_{t}$ with respect to $Q_{s}$ and $P_{t}$ governs which factorization scheme is relevant. Indeed, when $k_{t}\sim Q_{s}\ll P_{t}$ (the quark and the antiquark are almost back-to-back), there are effectively two strongly ordered scales $k_{t}$ and $P_{t}$ in the problem and TMD factorization applies \cite{Dominguez:2011wm}, implying the involvement of several gluon TMDs that differ significantly from each other, especially in the saturation regime, when $k_t\leq Q_s$ \cite{Marquet:2016cgx}. In the other regime: $Q_{s}\ll k_{t}\sim P_{t}$, $k_{t}$ and $P_{t}$ are of the same order and far above the saturation scale, hence high-energy factorization \cite{Catani:1990xk,Catani:1990eg} is applicable. In this case, only the linear small-$x$ dynamics governed by the Balitsky-Fadin-Kuraev-Lipatov (BFKL) equation \cite{Lipatov:1976zz,Kuraev:1976ge,Balitsky:1978ic} is important, and the TMDs differ no more, implying that only one such distribution plays a role. Interestingly, both regimes, i.e. the TMD regime and the high-energy factorization regime, are encompassed within the CGC approach \cite{Kotko:2015ura}.

Indeed, in \cite{Marquet:2007vb,Dominguez:2011wm}, the cross section for forward di-jet production in proton-nucleus collisions
was calculated within the CGC. It was then shown that, in the back-to-back limit $k_{t}\sim Q_{s}\ll P_{t}$, a TMD factorization formula could
be extracted, the result being the same as in a direct TMD approach (i.e., without resorting to the CGC). However, in contrast to the
direct TMD approach, the calculation in the CGC yields explicit expressions for the TMDs in terms of Wilson lines, which can be evolved in rapidity
through the nonlinear Jalilian-Marian-Iancu-McLerran-Weigert-Leonidov-Kovner (JIMWLK) equation, as was demonstrated in \cite{Marquet:2016cgx}.

In this paper, we build further on that work, by studying the forward production of a heavy quark-antiquark pair.
As already observed earlier (see for instance \cite{Mulders:2000sh,Boer:2009nc,Boer:2010zf,Metz:2011wb,dominguez:2011br,Akcakaya:2012si}),
by keeping a non-zero quark mass, the cross section becomes sensitive to additional TMDs, which describe the linearly-polarized gluon content
of the unpolarized target, or in our case, nucleus. The three unpolarized gluon TMDs that describe the gluon channel $gA\rightarrow q\bar{q}$ will
be accompanied by three \textquoteleft polarized' partners, which couple through the quark mass and via a $\mathrm{cos}(2\phi)$ modulation, where $\phi$ relates to the quark-antiquark pair and is defined below. This is analogous to what happens in the $\gamma^*A\rightarrow q\bar{q}$ process (in that case not only a non-zero quark mass but also a non-zero photon virtuality brings sensitivity to linearly-polarized gluons), although there only one unpolarized gluon TMDs is involved (the Weizsäcker-Williams distribution), along with its polarized partner \cite{Metz:2011wb,dominguez:2011br}.

The paper is organized as follows. In section II, we give the result of the CGC calculation of the forward heavy quark-antiquark pair production cross section, and demonstrate how the six gluon TMDs (three unpolarized and three linearly-polarized) emerge in the appropriate limit. In section III, we compute these TMDs analytically in the McLerran-Venugopalan (MV) model and compare our results with the existing literature, after which in section IV they are numerically evaluated and evolved in rapidity with the help of a lattice implementation of the JIMWLK equation. Finally, we conclude and give an outlook for further work.

\section{Extracting a TMD factorization formula from the CGC framework}
\label{sec:correlationlimit}

We consider inclusive quark-antiquark pair production in the forward
region, in collisions of dilute and dense systems 
\begin{equation}
p(p_{p})+A(p_{A})\to Q(p_{1})+\bar{Q}(p_{2})+X\ .
\end{equation}
The four-momenta of the projectile and the target are massless and
purely longitudinal. In terms of the light-cone variables, $x^{\pm}=(x^{0}\pm x^{3})/\sqrt{2}$,
they take the simple form $p_{p}=\sqrt{s/2}\ (1,0_{t},0)$ and $p_{A}=\sqrt{s/2}\ (0,0_{t},1)$,
where $s$ is the squared center of mass energy of the p+A system.
The energy (or longitudinal momenta) fractions $x_{1}$ and $x_{2}$
of the incoming gluons from the projectile and the target, respectively,
can be expressed in terms of the rapidities $(y_{1},y_{2})$ and transverse
momenta $(p_{1t},p_{2t})$ of the produced particles as 
\begin{equation}
\begin{aligned}x_{1} & =\frac{p_{1}^{+}+p_{2}^{+}}{p_{p}^{+}}=\frac{1}{\sqrt{s}}\left(\sqrt{p_{1t}^{2}+m^{2}}e^{y_{1}}+\sqrt{p_{2t}^{2}+m^{2}}e^{y_{2}}\right)\;,\\
x_{2} & =\frac{p_{1}^{-}+p_{2}^{-}}{p_{A}^{-}}=\frac{1}{\sqrt{s}}\left(\sqrt{p_{1t}^{2}+m^{2}}e^{-y_{1}}+\sqrt{p_{2t}^{2}+m^{2}}e^{-y_{2}}\right)\;,
\end{aligned}
\end{equation}
 where $m$ denotes the quark mass.

By imposing production in the forward direction, we effectively select
these fractions to be $x_{1}\sim1$ and $x_{2}\ll1$. Therefore, the
large-$x$ gluons of the dilute projectile are described in terms
of the usual gluon distribution of collinear factorization $g(x_{1},\mu^{2})$,
and the $pA\to Q\bar{Q}X$ cross section is obtained from the $gA\to Q\bar{Q}X$
cross section as:
\begin{equation}
\frac{\mathrm{d}\sigma(pA\to Q\bar{Q}X)}{\mathrm{d}^{3}p_{1}\mathrm{d}^{3}p_{2}}=\int \mathrm{d}x\ g(x,\mu^{2})\frac{\mathrm{d}\sigma(gA\to Q\bar{Q}X)}{\mathrm{d}^{3}p_{1}\mathrm{d}^{3}p_{2}}(p^{+}=xp_{p}^{+},p_{t}=0)\:,
\end{equation}
where $p=(p^{+},p_{t})$ denotes the momentum of the incoming gluon.

By contrast, due to the large gluon density of the small-$x_2$ gluons, the $gA\to Q\bar{Q}X$ cross section does not generally factorize further (it does if non-linear effects can be neglected): $\mathrm{d}\sigma(gA\to Q\bar{Q}X)\neq \mathrm{d}\sigma(gg\to Q\bar{Q}X) \otimes g_A $. This is due to the fact that in the saturation regime, the gluons in the nuclear wave function interact with the projectile in a coherent manner. Such density effects can be taken into account using the CGC description of the dense small-$x_2$ gluon content of the nucleus in terms of strong classical fields. Then, the $gA\to Q\bar{Q}X$ cross section involves averages over color field configurations which may be written as
\begin{equation}
\left\langle O\right\rangle _{x_{2}}=\int DA^{-}\mathcal{W}_{x_{2}}[A^{-}]O[A^{-}]\ ,
\end{equation}
where $\mathcal{W}_{x_{2}}[A^{-}]$ represents the probability of a given field configuration (we use a gauge in which $A^-$ is the only non zero component of the field). Let us now detail what these CGC averages exactly look like, and how an effective factorization with several TMDs emerges in the appropriate limit \cite{Taels:2017shj}.

\subsection{Starting CGC formulation}

\begin{figure}
\begin{centering}
\includegraphics[clip,scale=0.6]{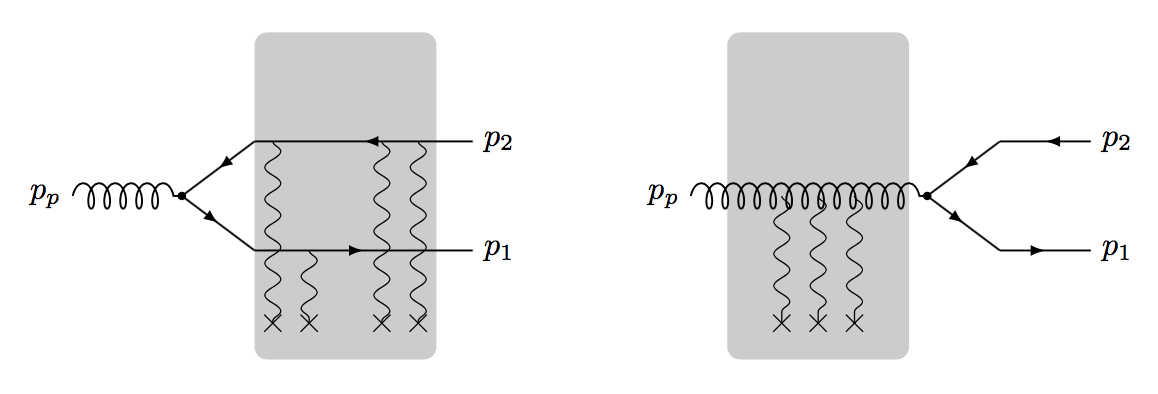}
\par\end{centering}
\caption{Amplitude for quark-antiquark production in the CGC formalism. The pair can be radiated before (left) or after (right) the interaction with the target. The two
terms come with a relative minus sign.}
\label{fig:amplitudeCGC} 
\end{figure}

Our starting point is the CGC formalism for quark-antiquark pair production
in dilute-dense collisions. The amplitude for quark-antiquark pair
production is schematically presented in Fig.~\ref{fig:amplitudeCGC}.
In the CGC formalism, the scattering of the partons from the dilute
projectile with the dense target is described by Wilson lines that
resum multi-gluon exchanges; fundamental Wilson lines for quarks and
adjoint Wilson lines for gluons. As a result, the cross section involves
multipoint correlators of Wilson lines. In particular, the square
of the amplitude from Fig.~\ref{fig:amplitudeCGC} contains four
terms: a correlator of four Wilson lines, $S^{(4)}$, corresponding
to interactions happening after the creation of the $q\bar{q}$ pair,
both in the amplitude and the complex conjugate, then a correlator
of two Wilson lines, $S^{(2)}$ representing the case when interactions
with the target take place before the gluon splits in both the amplitude
and the complex conjugate one, and two correlators of three Wilson
lines, $S^{(3)}$, for the cross terms.

Introducing 
\begin{equation}
z=\frac{p_{1}^{+}}{p_{1}^{+}+p_{2}^{+}}\ ,\quad\quad k_{t}=p_{1t}+p_{2t}\ ,\quad\quad\text{and}\quad\quad P_{t}=(1-z)p_{1t}-zp_{2t}\ ,
\end{equation}
the cross section reads \cite{Dominguez:2011wm}: 
\begin{eqnarray}
\frac{\mathrm{d}\sigma(pA\to Q(p_{1})\bar{Q}(p_{2})X)}{\mathrm{d}y_{1}\mathrm{d}y_{2}\mathrm{d}^{2}p_{1t}\mathrm{d}^{2}p_{2t}}=\frac{\alpha_{s}}{2}z(1-z)x_{1}g(x_{1},\mu^{2})\int\frac{\mathrm{d}^{2}{\bf u}}{(2\pi)^{2}}\frac{\mathrm{d}^{2}{\bf u'}}{(2\pi)^{2}}e^{i\mathbf{P}_{t}\cdot({\bf u'}-{\bf u})}\ p^{+}\!\sum_{\lambda\alpha\beta}\varphi_{\alpha\beta}^{\lambda^{*}}(p,p_{1}^{+},{\bf u'})\varphi_{\alpha\beta}^{\lambda}(p,p_{1}^{+},{\bf u})\hspace{1cm}\nonumber \\
\int\frac{\mathrm{d}^{2}{\bf v}}{(2\pi)^{2}}\frac{\mathrm{d}^{2}{\bf v'}}{(2\pi)^{2}}e^{i\mathbf{k}_{t}\cdot({\bf v'}-{\bf v})}\left\{ S_{q\bar{q}\bar{q}q}^{(4)}\left({\bf x},{\bf b},{\bf x'},{\bf b'};x_{2}\right)-S_{qg\bar{q}}^{(3)}\left({\bf x},{\bf v'},{\bf b};x_{2}\right)-S_{qg\bar{q}}^{(3)}\left({\bf b'},{\bf v},{\bf x'},x_{2}\right)+S_{gg}^{(2)}\left({\bf v},{\bf v'};x_{2}\right)\right\} \ , \hspace{0.3cm}\label{eq:cgc-qqbar}
\end{eqnarray}
where 
\begin{equation}
{\bf x}={\bf v}+(1\!-\!z){\bf u}\quad\mbox{and}\quad{\bf x'}={\bf v'}+(1\!-\!z){\bf u'}
\end{equation}
denote the transverse positions of the final-state quark in the amplitude
and the conjugate amplitude, respectively, and 
\begin{equation}
{\bf b}={\bf v}-z{\bf u}\quad\mbox{and}\quad{\bf b'}={\bf v'}-z{\bf u'}
\end{equation}
denote the transverse positions of the final-state antiquark in the
amplitude and the conjugate amplitude, respectively. The difference ${\bf {u'}}-{\bf {u}}$
is conjugate to the hard momentum $P_{t}$, and ${\bf {v'}}-{\bf {v}}$
is conjugate to the total transverse momentum of the pair $k_{t}$.

The $S^{(i)}$ Wilson line correlators are given by: 
\begin{eqnarray}
S_{q\bar{q}\bar{q}q}^{(4)}({\bf x},{\bf b},{\bf x'},{\bf b'};x_{2}) & = & \frac{1}{C_{F}N_{c}}\left<\text{Tr}\left(U_{{\bf b}}^{\dagger}t^{c}U_{{\bf x}}U_{{\bf x'}}^{\dagger}t^{c}U_{{\bf b'}}\right)\right>_{x_{2}}\ ,\\
S_{qg\bar{q}}^{(3)}({\bf x},{\bf v},{\bf b};x_{2}) & = & \frac{1}{C_{F}N_{c}}\left<\text{Tr}\left(U_{{\bf b}}^{\dagger}t^{c}U_{{\bf x}}t^{d}\right)V_{{\bf v}}^{cd}\right>_{x_{2}}\ ,\\
\label{eq:adjointdipole} S_{gg}^{(2)}({\bf v},{\bf v'};x_{2}) & = & \frac{1}{N_{c}^{2}-1}\left<\text{Tr}\left(V_{{\bf v}}V_{{\bf v'}}^{\dagger}\right)\right>_{x_{2}}\ ,
\end{eqnarray}
where 
\begin{equation}
U_{{\bf x}}=\mathcal{P}\exp\left[ig_s\int_{-\infty}^{\infty}\mathrm{d}x^{+}A_{a}^{-}(x^{+},{\bf x})t^{a}\right]\;,\quad\quad V_{{\bf x}}=\mathcal{P}\exp\left[ig_s\int_{-\infty}^{\infty}\mathrm{d}x^{+}A_{a}^{-}(x^{+},{\bf x})T^{a}\right]
\end{equation}
with $t^{a}$ and $T^{a}$ denoting the generators of the fundamental
and adjoint representation of $SU(N_{c})$, respectively.

The functions $\varphi_{\alpha\beta}^{\lambda}$ are the $g\to Q\bar{Q}$
splitting wave functions, and their overlap is given by: 
\begin{equation}
p^{+}\sum_{\lambda\alpha\beta}\varphi_{\alpha\beta}^{\lambda^{*}}(p,p_{1}^{+},{\bf {u'}})\varphi_{\alpha\beta}^{\lambda}(p,p_{1}^{+},{\bf {u}})=8\pi^{2}\left[2P_{qg}(z)\frac{{\bf {u}}\cdot{\bf {u'}}}{|{\bf {u}}||{\bf {u'}}|}m^{2}K_{1}(m|{\bf {u}}|)K_{1}(m|{\bf {u'}}|)+m^{2}K_{0}(m|{\bf {u}}|)K_{0}(m|{\bf {u'}}|)\right]\ ,\label{eq:overlap}
\end{equation}
with $m$ denoting the mass of the quark and with 
\begin{equation}
P_{qg}(z)=\frac{z^{2}+(1\!-\!z)^{2}}{2}\ .
\end{equation}

\subsection{Extracting the leading power}

In order to investigate the TMD regime, we shall extract the leading
power in $1/P_{t}^{2}$. This corresponds the quark and the antiquark
being emitted nearly back-to-back in the transverse plane, as the
total transverse momentum of the pair $|k_{t}|$ is required to be
much smaller than the individual transverse momenta. Importantly,
even though $Q_{s}^{2}$ is also required to be much smaller than
$P_{t}^{2}$, saturation effects still play a role, when $k_{t}^{2}\sim Q_{s}^{2}$.
In the case of massless quarks, the calculation was performed in \cite{Dominguez:2011wm}
in the large-$N_{c}$ limit and in \cite{Marquet:2016cgx} keeping
$N_{c}$ finite.

In the $|k_{t}|,Q_{s}\ll|P_{t}|$ limit, the integrals in \eqref{eq:cgc-qqbar}
are controlled by configurations where $|{\bf u}|$ and $|{\bf u'}|$
are small compared to the other transverse-size variables, and the
leading $1/P_{t}^{2}$ power of this expression can be extracted by
expanding around ${\bf b}={\bf x}={\bf v}$ and ${\bf b'}={\bf x'}={\bf v'}$.
To do this, let us first rewrite all the Wilson line correlators in
terms of fundamental Wilson lines only: 
\begin{eqnarray}
S_{q\bar{q}\bar{q}q}^{(4)}({\bf {x}},{\bf {b}},{\bf {x'}},{\bf {b'}};x_{2}) & = & \frac{N_{c}}{2C_{F}}\left<D({\bf x},{\bf x'})D({\bf b'},{\bf b})-\frac{1}{N_{c}^{2}}Q({\bf x},{\bf x'},{\bf b'},{\bf b})\right>_{x_{2}}\ ,\\
S_{qg\bar{q}}^{(3)}({\bf {x}},{\bf {v}},{\bf {b}};x_{2}) & = & \frac{N_{c}}{2C_{F}}\left<D({\bf x},{\bf v})D({\bf v},{\bf b})-\frac{1}{N_{c}^{2}}D({\bf x},{\bf b})\right>_{x_{2}}\ ,\\
S_{gg}^{(2)}({\bf {v}},{\bf {v'}};x_{2}) & = & \frac{N_{c}}{2C_{F}}\left<D({\bf v},{\bf v'})D({\bf v'},{\bf v})-\frac{1}{N_{c}^{2}}\right>_{x_{2}}\ ,
\end{eqnarray}
where 
\begin{equation}
D({\bf x},{\bf y})=\frac{1}{N_{c}}{\text{Tr}}\left(U_{{\bf x}}U_{{\bf y}}^{\dagger}\right)\quad\mbox{and}\quad Q({\bf x},{\bf y},{\bf v},{\bf w})=\frac{1}{N_{c}}{\text{Tr}}\left(U_{{\bf x}}U_{{\bf y}}^{\dagger}U_{{\bf v}}U_{{\bf w}}^{\dagger}\right)\ .
\label{eq:dipquad}
\end{equation}
Then, the combination inside the brackets $\Big\{.\Big\}$ in Eq.~\eqref{eq:cgc-qqbar}
can be rewritten:
\begin{equation}
\begin{aligned} & \frac{N_{c}}{2C_{F}}\Big<D[{\bf v}\!+\!(1\!-\!z){\bf u},{\bf v'}\!+\!(1\!-\!z){\bf u'}]D[{\bf v'}\!-\!z{\bf u'},{\bf v}\!-\!z{\bf u}]+D[{\bf v},{\bf v'}]D[{\bf v'},{\bf v}]\\
 & -D[{\bf v}\!+\!(1\!-\!z){\bf u},{\bf v'}]D[{\bf v'},{\bf v}\!-\!z{\bf u}]-D[{\bf v'}\!-\!z{\bf u'},{\bf v}]D[{\bf v},{\bf v'}\!+\!(1\!-\!z){\bf u'}]\Big>_{x_{2}}\\
 & -\frac{1}{2C_{F}N_{c}}\Big<1+Q[{\bf v}\!+\!(1\!-\!z){\bf u},{\bf v'}\!+\!(1\!-\!z){\bf u'},{\bf v'}\!-\!z{\bf u'},{\bf v}\!-\!z{\bf u}]\\
 & -D[{\bf v}\!+\!(1\!-\!z){\bf u},{\bf v}\!-\!z{\bf u}]-D[{\bf v'}\!-\!z{\bf u'},{\bf v'}\!+\!(1\!-\!z){\bf u'}]\Big>_{x_{2}}\ .
\end{aligned}
\label{eq:lt-qqbar}
\end{equation}
This expression vanishes if either \textbf{u} or ${\bf u'}$ is set
to zero. Therefore, the first non-zero term in its expansion is the
one that contains both one power of ${\bf u}$ and one power of ${\bf u'}$:
\begin{equation}
\begin{aligned} & \frac{N_{c}u^{i}u'^{j}}{2C_{F}}[(1\!-\!z)\partial_{v}^{i}-z\partial_{x}^{i}][(1\!-\!z)\partial_{v'}^{j}-z\partial_{y}^{j}]\Big<D({\bf v},{\bf v'})D({\bf y},{\bf x})-\frac{1}{N_{c}^{2}}Q({\bf v},{\bf v'},{\bf y},{\bf x})\Big>_{x_{2}}\Big|_{\substack{{\bf x}={\bf v}\\
{\bf y}={\bf v'}
}
}\\
 & =\frac{N_{c}u^{i}u'^{j}}{2C_{F}}\left[z^{2}\Big<D({\bf v},{\bf v'})\partial_{v}^{i}\partial_{v'}^{j}D({\bf v'},{\bf v})\Big>_{x_{2}}+(1-z)^{2}\Big<D({\bf v},{\bf v'})\partial_{v}^{i}\partial_{v'}^{j}D({\bf v'},{\bf v})\Big>_{x_{2}}^{*}\right.\\
 & \left.-2z(1-z)\mbox{Re}\Big<\left[\partial_{v}^{i}D({\bf v},{\bf v'})\right]\partial_{v'}^{j}D({\bf v'},{\bf v})\Big>_{x_{2}}+\frac{1}{N_{c}^{2}}\Big<\partial_{v}^{i}\partial_{y}^{j}Q({\bf v},{\bf v'},{\bf y},{\bf x})\Big>_{x_{2}}\Big|_{\substack{{\bf x}={\bf v}\\
{\bf y}={\bf v'}
}
}\right]\ .
\end{aligned}
\label{eq:expansion}
\end{equation}

So far the derivation has been identical to that of the massless case
in \cite{Marquet:2016cgx}, the difference resides in the wave function
overlap \eqref{eq:overlap}, which, after multiplication by $u^{i}u'^{j}$
and Fourier transformation, yields: 
\begin{equation}
\int\frac{\mathrm{d}^{2}{\bf u}}{(2\pi)^{2}}\frac{\mathrm{d}^{2}{\bf u'}}{(2\pi)^{2}}e^{i\mathbf{P}_{t}\cdot({\bf u'}-{\bf u})}u^{i}u'^{j}\ p^{+}\!\sum_{\lambda\alpha\beta}\varphi_{\alpha\beta}^{\lambda^{*}}(p,p_{1}^{+},{\bf u'})\varphi_{\alpha\beta}^{\lambda}(p,p_{1}^{+},{\bf u})=\frac{4P_{qg}(z)}{(P_{t}^{2}+m^{2})^{2}}\lr{\delta_{ij}-\frac{4m^{2}P_{i}P_{j}}{(P_{t}^{2}+m^{2})^{2}}}+\frac{8m^{2}P_{i}P_{j}}{(P_{t}^{2}+m^{2})^{4}}\ .
\end{equation}
In the massless case, the transverse indices of the various structures
in \eqref{eq:expansion} were projected onto $\delta_{ij}$ only,
and unpolarized gluon TMDs were emerging. Now the presence of the
mass is responsible for the appearance of new objects: the so-called
linearly-polarized gluon TMDs.

\subsection{Unpolarized and linearly-polarized gluon TMDs}

The last integrations which remain to be done correspond to definitions
of various gluon TMDs: 
\begin{eqnarray}
 \frac{4}{g_{s}^{2}N_c}\!\int\!\frac{\mathrm{d}^{2}{\bf v}\mathrm{d}^{2}{\bf v'}}{(2\pi)^{3}}e^{-i\mathbf{k}_{t}\cdot({\bf v}-{\bf v'})}\left\langle \text{Tr}\left[(\partial_{i}U_{{\bf v'}})(\partial_{j}U_{{\bf v}}^{\dagger})\right]\text{Tr}\left[U_{{\bf v}}U_{{\bf v'}}^{\dagger}\right]\right\rangle _{x_{2}}
\!=\!\frac{\delta_{ij}}{2}\mathcal{F}_{gg}^{(1)}(x_{2},k_{t})\!+\!\left(\frac{k_{i}k_{j}}{k_{t}^{2}}\!-\!\frac{\delta_{ij}}{2}\right)\!\mathcal{H}_{gg}^{(1)}(x_{2},k_{t})\ ,\label{eq:FH1inverted}\\\hspace{-0.5cm}\frac{-4}{g_{s}^{2}N_c}\!\int\!\frac{\mathrm{d}^{2}{\bf v}\mathrm{d}^{2}{\bf v'}}{(2\pi)^{3}}e^{-i\mathbf{k}_{t}\cdot({\bf v}-{\bf v'})}\mbox{Re}\left\langle \text{Tr}\left[(\partial_{i}U_{{\bf v}})U_{{\bf v'}}^{\dagger}\right]\!\text{Tr}\left[(\partial_{j}U_{{\bf v'}})U_{{\bf v}}^{\dagger}\right]\right\rangle _{x_{2}}
\!=\!\frac{\delta_{ij}}{2}\mathcal{F}_{gg}^{(2)}(x_{2},k_{t})\!+\!\left(\frac{k_{i}k_{j}}{k_{t}^{2}}\!-\!\frac{\delta_{ij}}{2}\right)\!\mathcal{H}_{gg}^{(2)}(x_{2},k_{t})\ ,\label{eq:FH2inverted}\\ \frac{-4}{g_{s}^{2}}\!\int\!\frac{\mathrm{d}^{2}{\bf v}\mathrm{d}^{2}{\bf v'}}{(2\pi)^{3}}e^{-i\mathbf{k}_{t}\cdot({\bf v}-{\bf v'})}\left\langle \text{Tr}\left[(\partial_{i}U_{{\bf v}})U_{{\bf v'}}^{\dagger}(\partial_{j}U_{{\bf v'}})U_{{\bf v}}^{\dagger}\right]\right\rangle _{x_{2}}
\!=\!\frac{\delta_{ij}}{2}\mathcal{F}_{gg}^{(3)}(x_{2},k_{t})\!+\!\left(\frac{k_{i}k_{j}}{k_{t}^{2}}\!-\!\frac{\delta_{ij}}{2}\right)\!\mathcal{H}_{gg}^{(3)}(x_{2},k_{t})\ .\label{eq:FH3inverted}
\end{eqnarray}
Both parts of the projection are gluon TMDs, as we will shortly demonstrate. Interestingly, the traceless parts -- $\mathcal{H}_{gg}^{\left(1,2,3\right)}$-- are the TMDs that correspond to the linearly polarized gluons inside the unpolarized nucleus \cite{Mulders:2000sh, Boer:2009nc,Metz:2011wb, dominguez:2011br}. Gluon polarization hence does play a role in forward heavy-quark production in dilute-dense collisions, even when those collisions involve unpolarized beams. The connection between the generic operator definitions of the gluon TMDs and the definitions given here, valid in the small-$x$ limit, was detailed in \cite{Marquet:2016cgx} (strictly speaking after projecting onto $\delta_{ij}$, but the derivation is identical otherwise), and a short summary can be found in appendix \ref{sec:operatordef}.

In the leading-logarithmic approximation, the evolution of the CGC wave function $\mathcal{W}_{x_{2}}[A^{-}]$ with decreasing $x_{2}$ is obtained from the JIMWLK equation $\mathrm{d}/\mathrm{d}\ln(1/x_{2})\mathcal{W}_{x_{2}}[A^{-}]=H_{JIMWLK}\mathcal{W}_{x_{2}}[A^{-}]$. In turn, CGC averages in general, and the 6 gluon TMDs introduced above in particular, also evolve towards small values of $x_{2}$ according to that non-linear equation. In addition, the scale dependence of those gluon TMDs (not made explicit here) can also be taken into account, although we leave for future work: at small-$x$ this boils down to implementing Sudakov factors into our formalism \cite{Mueller:2012uf,Mueller:2013wwa}.

\begin{figure}
\begin{centering}
\includegraphics[clip,scale=1.0]{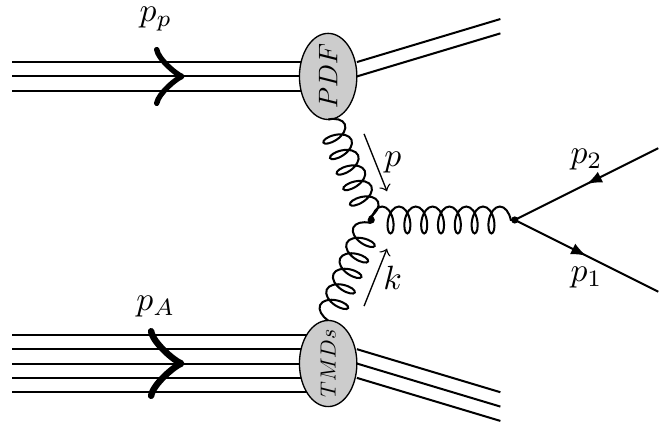}
\par\end{centering}
\caption{One of the leading order diagrams for inclusive heavy-quark pair production in p+A collisions.}
\label{fig:TMDfact} 
\end{figure}

Introducing the angle $\phi$ between $P_{t}$ and $k_{t}$, one can
write 
\begin{equation}
P_{i}P_{j}\left(\frac{k_{i}k_{j}}{k_{t}^{2}}-\frac{\delta_{ij}}{2}\right)=\frac{P_{t}^{2}}{2}\cos\left(2\phi\right)\ ,
\end{equation}
and put all the pieces together to finally obtain: 
\begin{eqnarray}
\frac{\mathrm{d}\sigma(pA\to Q(p_{1})\bar{Q}(p_{2})X)}{\mathrm{d}y_{1}\mathrm{d}y_{2}\mathrm{d}^{2}p_{1t}\mathrm{d}^{2}p_{2t}} & = & \frac{\alpha_{s}^{2}}{2C_{F}}\frac{z(1-z)}{(P_{t}^{2}+m^{2})^{2}}x_{1}g(x_{1},\mu^{2})\left\{ \lr{P_{qg}(z)+z(1-z)\frac{2m^{2}P_{t}^{2}}{(P_{t}^{2}+m^{2})^{2}}}\hspace{2.5cm}\nonumber\right.\\
 & \times & \lr{[(1\!-\!z)^{2}+z^{2}]{\cal F}_{gg}^{(1)}(x_{2},k_{t})+2z(1-z){\cal F}_{gg}^{(2)}(x_{2},k_{t})-\frac{1}{N_{c}^{2}}{\cal F}_{gg}^{(3)}(x_{2},k_{t})}\nonumber \\
 & + & z(1-z)\frac{2m^{2}P_{t}^{2}}{(P_{t}^{2}+m^{2})^{2}}\cos(2\phi)\nonumber \\
 & \times & \left.\lr{[(1\!-\!z)^{2}+z^{2}]{\cal H}_{gg}^{(1)}(x_{2},k_{t})+2z(1-z){\cal H}_{gg}^{(2)}(x_{2},k_{t})-\frac{1}{N_{c}^{2}}{\cal H}_{gg}^{(3)}(x_{2},k_{t})}\right\} \ .\label{eq:finalCGCcrosssection}
\end{eqnarray}
Therefore, the leading power of the CGC expression can be interpreted as a TMD factorization formula, with the small-$x_{2}$ gluon carrying a transverse momentum equal to $k_{t}$, and with several gluon TMDs needed to consistently describe the dense gluon content of the nucleus. This is illustrated by Figure \ref{fig:TMDfact}.

As is clear from the above formula, the information on the gluon polarization, encoded in $\mathcal{H}_{gg}^{\left(1,2,3\right)}$, couples to the
mass $m$ of the heavy quarks, and exhibits an angular dependence $\cos\left(2\phi\right)$, where $\phi$ is the angle between the
transverse momentum of one of the jets, and the transverse-momentum imbalance of the two jets.

\subsection{Final formula}

It is worth noting that $\mathcal{F}_{gg}^{\left(1\right)}$, $\mathcal{F}_{gg}^{\left(2\right)}$,
$\mathcal{H}_{gg}^{\left(1\right)}$, and $\mathcal{H}_{gg}^{\left(2\right)}$
are not independent, but instead are related to each other through
the dipole distribution in the adjoint representation: ${\cal F}_{ADP}(x_{2},k_{t})$
(different from the fundamental dipole gluon TMD ${\cal F}_{qg}^{(1)}={\cal F}_{DP}$),
defined as:
\begin{eqnarray}
{\cal F}_{ADP}(x_{2},k_{t}) & \equiv & \frac{4C_{F}k_{t}^{2}}{g_{s}^{2}}\int\frac{\mathrm{d}^{2}{\bf v}\mathrm{d}^{2}{\bf v'}}{(2\pi)^{3}}e^{-i\mathbf{k}_{t}\cdot({\bf v}-{\bf v'})}S_{gg}^{(2)}({\bf v},{\bf v'};x_{2})\;.
\end{eqnarray}

Indeed, we have:
\begin{eqnarray}
{\cal F}_{gg}^{(1)}(x_{2},k_{t})-{\cal F}_{gg}^{(2)}(x_{2},k_{t}) & = & \frac{2}{g_{s}^{2}N_{c}}\int\frac{\mathrm{d}^{2}{\bf v}\mathrm{d}^{2}{\bf v'}}{(2\pi)^{3}}e^{-i\mathbf{k}_{t}\cdot({\bf v}-{\bf v'})}\nabla_{{\bf v}}\cdot\nabla_{{\bf v'}}\left\langle N_{c}^{2}|D({\bf v},{\bf v'})|^{2}-1\right\rangle _{x_{2}}\;,\nonumber \\
 & = & \frac{2}{g_{s}^{2}N_{c}}\int\frac{\mathrm{d}^{2}{\bf v}\mathrm{d}^{2}{\bf v'}}{(2\pi)^{3}}e^{-i\mathbf{k}_{t}\cdot({\bf v}-{\bf v'})}\nabla_{{\bf v}}\cdot\nabla_{{\bf v'}}\left\langle \text{Tr}\left[V_{{\bf v}}V_{{\bf v'}}^{\dagger}\right]\right\rangle _{x_{2}}\;,\label{eq:F1minusF2}\\
 & = & {\cal F}_{ADP}(x_{2},k_{t})\;,\nonumber 
\end{eqnarray}
as well as:
\begin{eqnarray}
{\cal H}_{gg}^{(1)}(x_{2},k_{t})-{\cal H}_{gg}^{(2)}(x_{2},k_{t}) & = & \frac{2}{g_{s}^{2}N_{c}}\left(\frac{2k_{i}k_{j}}{k_{t}^{2}}-\delta_{ij}\right)\int\frac{\mathrm{d}^{2}{\bf v}\mathrm{d}^{2}{\bf v'}}{(2\pi)^{3}}e^{-i\mathbf{k}_{t}\cdot({\bf v}-{\bf v'})}\left\langle \text{Tr}\left[(\partial_{i}V_{{\bf v}})\partial_{j}V_{{\bf v'}}^{\dagger}\right]\right\rangle _{x_{2}}\;,\label{eq:H1minusH2}\\
 & = & {\cal F}_{ADP}(x_{2},k_{t})\;.\nonumber 
\end{eqnarray}

Therefore, the cross section may finally be written as:
\begin{eqnarray}
\frac{\mathrm{d}\sigma(pA\to Q(p_{1})\bar{Q}(p_{2})X)}{\mathrm{d}y_{1}\mathrm{d}y_{2}\mathrm{d}^{2}p_{1t}\mathrm{d}^{2}p_{2t}} & = & \frac{\alpha_{s}^{2}}{2C_{F}}\frac{z(1-z)}{(P_{t}^{2}+m^{2})^{2}}x_{1}g(x_{1},\mu^{2})\left\{ \lr{P_{qg}(z)+z(1-z)\frac{2m^{2}P_{t}^{2}}{(P_{t}^{2}+m^{2})^{2}}}\hspace{2.5cm}\nonumber\right.\\
 & \times & \lr{{\cal F}_{gg}^{(1)}(x_{2},k_{t})-2z(1-z){\cal F}_{ADP}(x_{2},k_{t})-\frac{1}{N_{c}^{2}}{\cal F}_{gg}^{(3)}(x_{2},k_{t})}\nonumber \\
 & + & z(1-z)\frac{2m^{2}P_{t}^{2}}{(P_{t}^{2}+m^{2})^{2}}\cos(2\phi)\nonumber \\
 & \times & \left.\lr{{\cal H}_{gg}^{(1)}(x_{2},k_{t})-2z(1-z){\cal F}_{ADP}(x_{2},k_{t})-\frac{1}{N_{c}^{2}}{\cal H}_{gg}^{(3)}(x_{2},k_{t})}\right\} \ .\label{eq:finalCGCcrosssectionDIPOLE}
\end{eqnarray}
This is our final formula. It is no more complicated than the one derived in \cite{Akcakaya:2012si} (3 Fs and 2 Hs) (which we show how to recover below), but it is more general (we did not assume the MV model and we kept the complete quark-mass dependence). Moreover, we clearly show that it is the adjoint-dipole gluon distribution which is involved. This TMD is different from the more familiar fundamental-dipole gluon TMD, but it features the same property that, in the small-$x$ limit, its unpolarized and linearly-polarized versions are identical.

In section \ref{sec:lattice}, we shall evaluate numerically all
the three unpolarized gluon TMDs 
\begin{eqnarray}
\mathcal{F}_{gg}^{(1)}(x_{2},k_{t}) & = & \frac{4}{g_{s}^{2}}\int\frac{\mathrm{d}^{2}{\bf v}\mathrm{d}^{2}{\bf v'}}{(2\pi)^{3}}e^{-i\mathbf{k}_{t}\cdot({\bf v}-{\bf v'})}\frac{1}{N_{c}}\left\langle \text{Tr}\left[(\partial_{i}U_{{\bf v'}})(\partial_{i}U_{{\bf v}}^{\dagger})\right]\text{Tr}\left[U_{{\bf v}}U_{{\bf v'}}^{\dagger}\right]\right\rangle _{x_{2}}\ ,\label{eq:F1}\\
\mathcal{F}_{gg}^{(2)}(x_{2},k_{t}) & = & -\frac{4}{g_{s}^{2}}\int\frac{\mathrm{d}^{2}{\bf v}\mathrm{d}^{2}{\bf v'}}{(2\pi)^{3}}e^{-i\mathbf{k}_{t}\cdot({\bf v}-{\bf v'})}\frac{1}{N_{c}}\mbox{Re}\left\langle \text{Tr}\left[(\partial_{i}U_{{\bf v}})U_{{\bf v'}}^{\dagger}\right]\text{Tr}\left[(\partial_{i}U_{{\bf v'}})U_{{\bf v}}^{\dagger}\right]\right\rangle _{x_{2}},\label{eq:F2}\\
\mathcal{F}_{gg}^{(3)}(x_{2},k_{t}) & = & -\frac{4}{g_{s}^{2}}\int\frac{\mathrm{d}^{2}{\bf v}\mathrm{d}^{2}{\bf v'}}{(2\pi)^{3}}e^{-i\mathbf{k}_{t}\cdot({\bf v}-{\bf v'})}\left\langle \text{Tr}\left[(\partial_{i}U_{{\bf v}})U_{{\bf v'}}^{\dagger}(\partial_{i}U_{{\bf v'}})U_{{\bf v}}^{\dagger}\right]\right\rangle _{x_{2}}\ ,\label{eq:F3}
\end{eqnarray}
as well as their linearly-polarized partners 
\begin{eqnarray}
\mathcal{H}_{gg}^{(1)}(x_{2},k_{t}) & = & \left(\frac{2k_{i}k_{j}}{k_{t}^{2}}-\delta_{ij}\right)\frac{4}{g_{s}^{2}}\int\frac{\mathrm{d}^{2}{\bf v}\mathrm{d}^{2}{\bf v'}}{(2\pi)^{3}}e^{-i\mathbf{k}_{t}\cdot({\bf v}-{\bf v'})}\frac{1}{N_{c}}\left\langle \text{Tr}\left[(\partial_{i}U_{{\bf v'}})(\partial_{j}U_{{\bf v}}^{\dagger})\right]\text{Tr}\left[U_{{\bf v}}U_{{\bf v'}}^{\dagger}\right]\right\rangle _{x_{2}}\ ,\label{eq:H1}\\
\mathcal{H}_{gg}^{(2)}(x_{2},k_{t}) & = & \left(\frac{2k_{i}k_{j}}{k_{t}^{2}}-\delta_{ij}\right)\lr{-\frac{4}{g_{s}^{2}}}\int\frac{\mathrm{d}^{2}{\bf v}\mathrm{d}^{2}{\bf v'}}{(2\pi)^{3}}e^{-i\mathbf{k}_{t}\cdot({\bf v}-{\bf v'})}\frac{1}{N_{c}}\mbox{Re}\left\langle \text{Tr}\left[(\partial_{i}U_{{\bf v}})U_{{\bf v'}}^{\dagger}\right]\text{Tr}\left[(\partial_{j}U_{{\bf v'}})U_{{\bf v}}^{\dagger}\right]\right\rangle _{x_{2}},\label{eq:H2}\\
\mathcal{H}_{gg}^{(3)}(x_{2},k_{t}) & = & \left(\frac{2k_{i}k_{j}}{k_{t}^{2}}-\delta_{ij}\right)\lr{-\frac{4}{g_{s}^{2}}}\int\frac{\mathrm{d}^{2}{\bf v}\mathrm{d}^{2}{\bf v'}}{(2\pi)^{3}}e^{-i\mathbf{k}_{t}\cdot({\bf v}-{\bf v'})}\left\langle \text{Tr}\left[(\partial_{i}U_{{\bf v}})U_{{\bf v'}}^{\dagger}(\partial_{j}U_{{\bf v'}})U_{{\bf v}}^{\dagger}\right]\right\rangle _{x_{2}}\ ,\label{eq:H3}
\end{eqnarray}
using a lattice calculation to solve the JIMWLK equation 
\begin{equation}
\frac{\mathrm{d}}{\mathrm{d}\mathrm{log}\left(1/x_{2}\right)}\langle O\rangle_{x_{2}}=\langle H_{JIMWLK}O\rangle_{x_{2}}\;.
\end{equation}
We will also evaluate ${\cal F}_{ADP}$ directly and check that $\mathcal{F}_{gg}^{(1)}-\mathcal{F}_{gg}^{(2)}=\mathcal{H}_{gg}^{(1)}-\mathcal{H}_{gg}^{(2)}=\mathcal{F}_{ADP}$.

\section{Analytical results in the MV model}
\label{sec:MV}

\subsection{McLerran-Venugopalan model}

The McLerran-Venugopalan (MV) model \cite{McLerran:1993ni,McLerran:1993ka,McLerran:1994vd} (see
also \cite{Iancu:2005jft}) is a classical model for the gluon
distribution in a large nucleus. It assumes a Gaussian distribution
of color charges, which act as static sources, generating
the soft gluons through the Yang-Mills equations. The two-point function of the gluon field $A_{a}^{-}$ is given by:
\begin{equation}
\begin{aligned}\langle A_{a}^{-}\left(x^{+},\mathbf{x}\right)A_{b}^{-}\left(y^{+},\mathbf{y}\right)\rangle_{A} & =\frac{1}{g_{s}^{2}}\delta_{ab}\delta\left(x^{+}-y^{+}\right)\lambda_{A}\left(x^{+}\right)L_{\mathbf{x}\mathbf{y}}\;,\end{aligned}
\end{equation}
with
\begin{equation}
\begin{aligned}L_{\mathbf{x}\mathbf{y}} & \equiv g_{s}^{2}\int\frac{\mathrm{d}^{2}\mathbf{q}_{t}}{\left(2\pi\right)^{2}}\frac{e^{i\mathbf{q}_{t}\cdot\left(\mathbf{x}-\mathbf{y}\right)}}{q_{t}^{4}}\;,\end{aligned}
\label{eq:L}
\end{equation}
where $\lambda_{A}\left(x^{+}\right)$ is the density of color
charge squared of the valence quarks, per unit volume and per color.
Its precise dependence on $x^{+}$ is not important, since all final
results only depend on the integrated density $\mu_{A}$, given by: 
\begin{equation}
\mu_{A}\equiv \int\mathrm{d}x^{+}\lambda_{A}\left(x^{+}\right)\label{eq:muA}\ .
\end{equation}

Evaluating the dipoles, defined in Eqs. \eqref{eq:adjointdipole} and \eqref{eq:dipquad}, within the MV model yields:
\begin{equation}
\begin{aligned}
S_{gg}^{(2)}({\bf x},{\bf y})=e^{-\frac{N_c}{2}\Gamma\left(\mathbf{x}-\mathbf{y}\right)}\quad\mbox{and}\quad
S_{q\bar q}^{(2)}({\bf x},{\bf y})=\left\bra D\left(\mathbf{x},\mathbf{y}\right)\right\ket_A =e^{-\frac{C_{F}}{2}\Gamma\left(\mathbf{x}-\mathbf{y}\right)}\;,\end{aligned}
\label{eq:DipoleMVGamma}
\end{equation}
where we introduced the dimensionless quantity:
\begin{equation}
\begin{aligned}\Gamma\left(\mathbf{x}-\mathbf{y}\right) & \equiv\mu_{A}\left(L_{\mathbf{xx}}+L_{\mathbf{yy}}-2L_{\mathbf{xy}}\right)=2\mu_{A}g_{s}^{2}\int\frac{\mathrm{d}^{2}\mathbf{q}_{t}}{\left(2\pi\right)^{2}}\frac{1}{q_{t}^{4}}\left(1-e^{i\mathbf{q}_{t}\cdot\left(\mathbf{x}-\mathbf{y}\right)}\right)\ .\end{aligned}
\end{equation}
After regulating the infrared, the integral above can be evaluated to logarithmic accuracy, giving:
\begin{equation}
\begin{aligned}
\Gamma\left(\mathbf{r}\right) \simeq \alpha_{s}\mu_{A}\frac{r^{2}}{2}\ln\frac{1}{r^{2}\Lambda^{2}} =\frac{r^{2}}{2}\frac{1}{C_{F}}Q_{s}^{2}\left(r\right)=\frac{r^{2}}{2}\frac{1}{N_{c}}Q_{sg}^{2}\left(r\right),
\end{aligned}
\label{eq:Gamma}
\end{equation}
where we have defined: 
\begin{equation}
\begin{aligned}Q_{s}^{2}\left(r\right) & \equiv\alpha_{s}C_{F}\mu_{A}\ln\frac{1}{r^{2}\Lambda^{2}}\;,\qquad Q_{sg}^{2}\left(r\right)\equiv\alpha_{s}N_{c}\mu_{A}\ln\frac{1}{r^{2}\Lambda^{2}}\;.\end{aligned}
\label{eq:SaturationScale}
\end{equation}
The above transverse momentum scales are the saturation scales experienced
by a quark or a gluon, respectively. These definitions allow to write 
\begin{equation}
\begin{aligned}
S_{q\bar q}^{(2)}(r)=e^{-\frac{r^{2}}{4}Q_{s}^{2}\left(r\right)}\quad\mbox{and}\quad S_{gg}^{(2)}(r)= e^{-\frac{r^{2}}{4}Q_{sg}^{2}\left(r\right)}\ .\end{aligned}
\label{eq:dipoleMVQs}
\end{equation}
Note that the MV model is purely classical, valid for a large nucleus and for values of $x_2$ of the order of $\sim10^{-2}$. Hence, at this point, there is no evolution, which is why the subscript $x_2$ in the target averages is omitted.

\subsection{Expressions for the gluon TMDs}

Let us start with the Weizsäcker-Williams gluon TMD $\mathcal{F}_{gg}^{\left(3\right)}$
and its partner $\mathcal{H}_{gg}^{\left(3\right)}$. As is clear
from their definition in Eqs. \eqref{eq:F3} and \eqref{eq:H3},
their main ingredient is a double derivative of the quadrupole correlator:
\begin{equation}
\left.\frac{\partial}{\partial x^{i}}\frac{\partial}{\partial y^{j}}\frac{1}{N_{c}}\left\langle \text{Tr}\left[U_{{\bf x}}U_{\mathbf{v}'}^{\dagger}U_{\mathbf{y}}U_{{\bf v}}^{\dagger}\right]\right\rangle _{x_{2}}\right|_{\mathbf{x}=\mathbf{v},\,\mathbf{y}=\mathbf{v}'}\;.\label{eq:derivquad}
\end{equation}
The expression for the quadrupole in the MV model was obtained in \cite{Dominguez:2011wm}, and reads: 
\begin{equation}
\begin{aligned} & \frac{1}{N_{c}}\text{Tr}\left\langle U_{{\bf x}}U_{\mathbf{v}'}^{\dagger}U_{\mathbf{y}}U_{{\bf v}}^{\dagger}\right\rangle _{x_{2}}\\
 & =e^{-\frac{C_{F}}{2}\left(\Gamma\left(\mathbf{x}-\mathbf{v}\right)+\Gamma\left(\mathbf{y}-\mathbf{v}'\right)\right)}e^{-\frac{N_{c}}{4}\mu_AF\left(\mathbf{x},\mathbf{y};\mathbf{v},\mathbf{v}'\right)+\frac{1}{2N_{c}}\mu_AF\left(\mathbf{x},\mathbf{v};\mathbf{y},\mathbf{v}'\right)}\\
 & \times\Biggl[\left(\frac{\sqrt{\Delta}+F\left(\mathbf{x},\mathbf{y};\mathbf{v},\mathbf{v}'\right)}{2\sqrt{\Delta}}-\frac{F\left(\mathbf{x},\mathbf{v};\mathbf{y},\mathbf{v}'\right)}{\sqrt{\Delta}}\right)e^{\frac{N_{c}}{4}\mu_A\sqrt{\Delta}}\\
 & +\left(\frac{\sqrt{\Delta}-F\left(\mathbf{x},\mathbf{y};\mathbf{v},\mathbf{v}'\right)}{2\sqrt{\Delta}}+\frac{F\left(\mathbf{x},\mathbf{v};\mathbf{y},\mathbf{v}'\right)}{\sqrt{\Delta}}\right)e^{-\frac{N_{c}}{4}\mu_A\sqrt{\Delta}}\Biggr]\;,
\end{aligned}
\label{eq:quadMV}
\end{equation}
with:
\begin{equation}
\begin{aligned}F\left(\mathbf{x},\mathbf{y},\mathbf{v},\mathbf{w}\right) & \equiv L_{\mathbf{xv}}-L_{\mathbf{xw}}+L_{\mathbf{yw}}-L_{\mathbf{yv}}\;,\end{aligned}
\label{eq:Fxyzw}
\end{equation}
and
\begin{equation}
\Delta\equiv F^{2}\left(\mathbf{x},\mathbf{y};\mathbf{\mathbf{v}},\mathbf{v}'\right)+\frac{4}{N_{c}^{2}}F\left(\mathbf{x},\mathbf{\mathbf{v}};\mathbf{y},\mathbf{v}'\right)F\left(\mathbf{x},\mathbf{v}';\mathbf{y},\mathbf{\mathbf{v}}\right)\;.
\end{equation}
Plugging expression Eq. \eqref{eq:quadMV} into \eqref{eq:derivquad}, one obtains:
\begin{equation}
\begin{aligned}\left.\frac{\partial}{\partial x^{i}}\frac{\partial}{\partial y^{j}}\frac{1}{N_{c}}\mathrm{Tr}\Bigl\langle U_{{\bf x}}U_{\mathbf{v}'}^{\dagger}U_{\mathbf{y}}U_{{\bf v}}^{\dagger}\Bigr \rangle_{x_{2}}\right|_{\mathbf{x}=\mathbf{v},\,\mathbf{y}=\mathbf{v}'} & =\frac{C_{F}}{N_{c}}\frac{1-e^{-\frac{N_{c}}{2}\Gamma\left(\mathbf{v}-\mathbf{v}'\right)}}{\Gamma\left(\mathbf{v}-\mathbf{v}'\right)}\frac{\partial}{\partial v^{i}}\frac{\partial}{\partial v'^{j}}\Gamma\left(\mathbf{v}-\mathbf{v}'\right)\;.\end{aligned}
\end{equation}
To proceed, we can evaluate the derivative of $\Gamma(\mathbf{v}-\mathbf{v}')$ further: 
\begin{equation}
\begin{aligned}\frac{\partial}{\partial v^{i}}\frac{\partial}{\partial v'^{j}}\Gamma\left(\mathbf{v}-\mathbf{v}'\right) & =-2g_{s}^{2}\mu_{A}\int\frac{\mathrm{d}^{2}\mathbf{l}}{\left(2\pi\right)^{2}}\frac{l^{i}l^{j}}{l^{4}}e^{i\mathbf{l}\cdot\left(\mathbf{v}-\mathbf{v}'\right)}\;,\end{aligned}
\label{eq:Dgamma}
\end{equation}
which, depending on the projection of the Lorentz indices, gives:
\begin{equation}
\begin{aligned}\delta^{ij}\frac{\partial}{\partial v^{i}}\frac{\partial}{\partial v'^{j}}\Gamma\left(\mathbf{v}-\mathbf{v}'\right) & =-\frac{g_{s}^{2}\mu_{A}}{\pi}\ln\frac{1}{\left|\mathbf{v}-\mathbf{v}'\right|\Lambda}\;,\end{aligned}
\label{eq:I1}
\end{equation}
or
\begin{equation}
\begin{aligned}\left(\frac{2k_{t}^{i}k_{t}^{j}}{k_{t}^{2}}-\delta^{ij}\right)\frac{\partial}{\partial v^{i}}\frac{\partial}{\partial v'^{j}}\Gamma\left(\mathbf{v}-\mathbf{v}'\right)
 & =-\frac{2g_{s}^{2}\mu_{A}}{\left(2\pi\right)^{2}}\int\mathrm{d}l\mathrm{d}\theta\frac{1}{l}e^{il\left|\mathbf{v}-\mathbf{v}'\right|\cos(\theta)}\left(2\cos^{2}\left(\theta+\alpha\right)-1\right)\;,\\
 & =\frac{g_{s}^{2}\mu_{A}}{\pi}\cos\left(2\alpha\right)\int_{0}^{\infty}\frac{\mathrm{d}l}{l}J_{2}\left(l\left|\mathbf{v}-\mathbf{v}'\right|\right)\;,\\ &=2\alpha_{s}\mu_{A}\cos\left(2\alpha\right)\;,
\end{aligned}
\label{eq:I2}
\end{equation}
where $\mathrm{cos}\, \alpha=\mathbf{k}_{t}\cdot \left(\mathbf{v}-\mathbf{v}'\right)/\left(|k_t|\cdot |\mathbf{v}-\mathbf{v}'|\right)$.
Using the integral representation of the Bessel function of the first kind:
\begin{equation}
\begin{aligned}\int_{0}^{2\pi}\mathrm{d}\phi\,e^{-ik_{t}r\cos(\phi)}\cos(2\phi) & =-2\pi J_{2}\left(k_{t}r\right)\;,\end{aligned}
\label{eq:J2}
\end{equation}
and the definition of the saturation scale in the MV model, Eq. \eqref{eq:SaturationScale},
one then finally obtains the following expressions for the Weizsäcker-Williams
gluon distribution and its polarized partner (in accordance
with the literature \cite{Dominguez:2010xd,Metz:2011wb,dominguez:2011br}):
\begin{equation}
\begin{aligned}\mathcal{F}_{gg}^{\left(3\right)}\left(x_{2},k_{t}\right) & =\frac{2C_{F}S_{\perp}}{\alpha_{s}\pi^{2}}\int\frac{\mathrm{d}^{2}\mathbf{r}}{\left(2\pi\right)^{2}}e^{-i\mathbf{k}_{t}\cdot\mathbf{r}}\frac{1}{r^{2}}\left(1-e^{-\frac{r^{2}}{4}Q_{sg}^{2}\left(r\right)}\right)\;,\end{aligned}
\label{eq:F3MV}
\end{equation}
\begin{equation}
\begin{aligned}\mathcal{H}_{gg}^{\left(3\right)}\left(x,q_{t}\right) & =\frac{C_{F}S_{\perp}}{\alpha_{s}\pi^{3}}\int\mathrm{d}r\frac{J_{2}\left(q_{t}r\right)}{r\ln\frac{1}{r^{2}\Lambda^{2}}}\left(1-e^{-\frac{r^{2}}{4}Q_{sg}^{2}\left(r\right)}\right)\;,\end{aligned}
\label{eq:H3MV}
\end{equation}
where $S_{\perp}$ denotes the transverse area of the nucleus:
\begin{equation}
S_{\perp}\equiv\int_{\mathrm{nucleus}}\mathrm{d}^{2}\mathbf{x}\;.
\end{equation}

The calculation of the four other gluon TMDs, $\mathcal{F}_{gg}^{(1)}$,
$\mathcal{F}_{gg}^{(2)}$, $\mathcal{H}_{gg}^{(1)}$ and $\mathcal{H}_{gg}^{(2)}$,
is analogous to the one above. Indeed, once again the main ingredient
of the gluon TMDs is a double derivative of a correlator of Wilson
lines. This time, it is the correlator of the product of two dipoles,
which was calculated in the MV model in \cite{Dominguez:2008aa}:
\begin{eqnarray}
\label{eq:dipdipMV}
\frac{1}{N_{c}^{2}}\Bigl\langle\mathrm{Tr}\left[U_{\mathbf{x}}U_{\mathbf{y}}^{\dagger}\right]\mathrm{Tr}\left[U_{\mathbf{v}'}U_{\mathbf{v}}^{\dagger}\right]\Bigr\rangle_{x_{2}} & = & e^{-\frac{C_{F}}{2}\left(\Gamma\left(\mathbf{x}-\mathbf{y}\right)+\Gamma\left(\mathbf{v}'-\mathbf{v}\right)\right)}e^{-\frac{N_{c}}{4}\mu_AF\left(\mathbf{x},\mathbf{v}';\mathbf{y},\mathbf{v}\right)+\frac{1}{2N_{c}}\mu_AF\left(\mathbf{x},\mathbf{y};\mathbf{v}',\mathbf{v}\right)}\nonumber \\
 &  & \times\Biggl[\left(\frac{F\left(\mathbf{x},\mathbf{v}';\mathbf{y},\mathbf{v}\right)+\sqrt{D}}{2\sqrt{D}}-\frac{F\left(\mathbf{x},\mathbf{y};\mathbf{v}',\mathbf{v}\right)}{N_{c}^{2}\sqrt{D}}\right)e^{\frac{N_{c}}{4}\mu_A\sqrt{D}}\\
 &  & -\left(\frac{F\left(\mathbf{x},\mathbf{v}';\mathbf{y},\mathbf{v}\right)-\sqrt{D}}{2\sqrt{D}}-\frac{F\left(\mathbf{x},\mathbf{y};\mathbf{v}',\mathbf{v}\right)}{N_{c}^{2}\sqrt{D}}\right)e^{-\frac{N_{c}}{4}\mu_A\sqrt{D}}\Biggr]\;,\nonumber 
\end{eqnarray}
where
\begin{equation}
D\equiv F^{2}\left(\mathbf{x},\mathbf{v}';\mathbf{y},\mathbf{v}\right)+\frac{4}{N_{c}^{2}}F\left(\mathbf{x},\mathbf{\mathbf{y}};\mathbf{v}',\mathbf{v}\right)F\left(\mathbf{x},\mathbf{v};\mathbf{v}',\mathbf{y}\right)\;,
\end{equation}
and with $F$ the same as in Eq. \eqref{eq:Fxyzw}. The gluon TMDs
$\mathcal{F}_{gg}^{(1)}$ and $\mathcal{H}_{gg}^{(1)}$, see Eqs. \eqref{eq:F1}, \eqref{eq:H1}, are built
from the following structure:
\begin{equation}
 \frac{1}{N_{c}^{2}}\left.\frac{\partial}{\partial x^{i}}\frac{\partial}{\partial y^{j}}\Bigl\langle\mathrm{Tr}\left[U_{\mathbf{x}}U_{\mathbf{y}}^{\dagger}\right]\mathrm{Tr}\left[U_{\mathbf{v}'}U_{\mathbf{v}}^{\dagger}\right]\Bigr\rangle_{x_{2}}\right|_{\mathbf{x}=\mathbf{v},\,\mathbf{y}=\mathbf{v}'}\;,
\end{equation}
which, with the help of Eq. \eqref{eq:dipdipMV}, becomes in the MV model:
\begin{equation}
\begin{aligned} & \frac{1}{N_{c}^{2}}\left.\frac{\partial}{\partial x^{i}}\frac{\partial}{\partial y^{j}}\Bigl\langle\mathrm{Tr}\left[U_{\mathbf{x}}U_{\mathbf{y}}^{\dagger}\right]\mathrm{Tr}\left[U_{\mathbf{v}'}U_{\mathbf{v}}^{\dagger}\right]\Bigr\rangle_{x_{2}}\right|_{\mathbf{x}=\mathbf{v},\,\mathbf{y}=\mathbf{v}'}\\
 & =\frac{C_{F}}{8N_{c}^{3}}\frac{e^{-\frac{N_{c}}{2}\Gamma\left(\mathbf{v}-\mathbf{v}'\right)}}{\Gamma\left(\mathbf{v}-\mathbf{v}'\right)}\Biggl[16\left(1-e^{\frac{N_{c}}{2}\Gamma\left(\mathbf{v}-\mathbf{v}'\right)}\right)\frac{\partial^{2}}{\partial v^{i}\partial v'^{j}}\Gamma\left(\mathbf{v}-\mathbf{v}'\right)\\
 & +\Gamma\left(\mathbf{v}-\mathbf{v}'\right)\Biggl(N_{c}^{4}\frac{\partial}{\partial x^{i}}\Gamma\left(\mathbf{x}-\mathbf{v}'\right)\frac{\partial}{\partial y^{j}}\Gamma\left(\mathbf{v}-\mathbf{y}\right) \bigg |_{\mathbf{x}=\mathbf{v},\mathbf{y}=\mathbf{v}'}  -4N_{c}\left(N_{c}^{2}-2\right)\frac{\partial^{2}}{\partial v^{i}\partial v'^{j}}\Gamma\left(\mathbf{v}-\mathbf{v}'\right)\Biggr)\Biggr]\;.
\end{aligned}
\label{eq:deriv2dipdip}
\end{equation}

Likewise, $\mathcal{F}_{gg}^{\left(2\right)}$ and $\mathcal{H}_{gg}^{\left(2\right)}$
are built from the structure:
\begin{equation}
\frac{1}{N_{c}^{2}}\frac{\partial^{2}}{\partial x^{i}\partial y^{j}}\left.\mathrm{Re}\Bigl\langle\mathrm{Tr}\left[U_{\mathbf{x}}U_{\mathbf{v}'}^{\dagger}\right]\mathrm{Tr}\left[U_{\mathbf{y}}U_{\mathbf{v}}^{\dagger}\right]\Bigr\rangle_{x_{2}}\right|_{\mathbf{x}=\mathbf{v},\mathbf{y}=\mathbf{v}'}\;,
\end{equation}
and one obtains:
\begin{equation}
\begin{aligned} & \frac{1}{N_{c}^{2}}\frac{\partial^{2}}{\partial x^{i}\partial y^{j}}\left.\mathrm{Re}\Bigl\langle\mathrm{Tr}\left[U_{\mathbf{x}}U_{\mathbf{v}'}^{\dagger}\right]\mathrm{Tr}\left[U_{\mathbf{y}}U_{\mathbf{v}}^{\dagger}\right]\Bigr\rangle_{x_{2}}\right|_{\mathbf{x}=\mathbf{v},\mathbf{y}=\mathbf{v}'}\\
 & =\frac{C_{F}}{8N_{c}^{3}}\frac{e^{-\frac{N_{c}}{2}\Gamma\left(\mathbf{v}-\mathbf{v}'\right)}}{\Gamma\left(\mathbf{v}-\mathbf{v}'\right)}\Biggl[-16\left(1-e^{\frac{N_{c}}{2}\Gamma\left(\mathbf{v}-\mathbf{v}'\right)}\right)\frac{\partial^{2}}{\partial v^{i}\partial v'^{j}}\Gamma\left(\mathbf{v}-\mathbf{v}'\right)\\
 & +\Gamma\left(\mathbf{v}-\mathbf{v}'\right)\left(N_{c}^{4}\left.\frac{\partial}{\partial x^{i}}\Gamma\left(\mathbf{x}-\mathbf{v}'\right)\frac{\partial}{\partial y^{j}}\Gamma\left(\mathbf{v}-\mathbf{y}\right)\right|_{\mathbf{x}=\mathbf{v},\mathbf{y}=\mathbf{v}'}-8N_{c}\frac{\partial^{2}}{\partial v^{i}\partial v'^{j}}\Gamma\left(\mathbf{v}-\mathbf{v}'\right)\right)\Biggr]\;.
\end{aligned}
\label{eq:dipdipderiv2}
\end{equation}

From these expressions, one can write
\begin{equation}
\eqref{eq:deriv2dipdip}+\eqref{eq:dipdipderiv2}=\frac{C_F}{N_c} \frac{\partial^{2}}{\partial v^{i}\partial v'^{j}} e^{-\frac{N_{c}}{2}\Gamma\left(\mathbf{v}-\mathbf{v}'\right)}
\end{equation}
which allows to further obtain $\mathcal{F}_{gg}^{(1)}-\mathcal{F}_{gg}^{(2)}=\mathcal{H}_{gg}^{(1)}-\mathcal{H}_{gg}^{(2)}=\mathcal{F}_{ADP}$, showing that these exact relations are not spoiled by the MV model assumptions. It is also possible to obtain more explicit expressions, using the following intermediate results:
\begin{equation}
\begin{aligned}\left.\delta^{ij}\frac{\partial}{\partial x^{i}}\Gamma\left(\mathbf{x}-\mathbf{v}'\right)\frac{\partial}{\partial y^{i}}\Gamma\left(\mathbf{v}-\mathbf{y}\right)\right|_{\mathbf{x}=\mathbf{v},\mathbf{y}=\mathbf{v}'} & =\left(2\mu_{A}g_{s}^{2}\right)^{2}\int\frac{\mathrm{d}^{2}\mathbf{k}}{\left(2\pi\right)^{2}}\int\frac{\mathrm{d}^{2}\mathbf{l}}{\left(2\pi\right)^{2}}\frac{\mathbf{k}\cdot\mathbf{l}}{l^{4}k^{4}}e^{i\left(\mathbf{k}+\mathbf{l}\right)\cdot\left(\mathbf{\mathbf{v}}-\mathbf{\mathbf{v}'}\right)}\;,\\
 & =\frac{\left(2\mu_{A}g_{s}^{2}\right)^{2}}{\left(2\pi\right)^{4}}\int\frac{\mathrm{d}k\mathrm{d}\theta}{k^{2}}\int\frac{\mathrm{d}l\mathrm{d}\phi}{l^{2}}\cos\left(\phi-\theta\right)e^{ik\left|\mathbf{\mathbf{v}}-\mathbf{\mathbf{v}'}\right|\cos\left(\theta\right)}e^{il\left|\mathbf{\mathbf{v}}-\mathbf{\mathbf{v}'}\right|\mathbf{\cos\left(\phi\right)}}\;,\\
 & =-\frac{\left(2\mu_{A}g_{s}^{2}\right)^{2}}{\left(2\pi\right)^{2}}\int_{\Lambda}^{\infty}\frac{\mathrm{d}k}{k^{2}}\int_{\Lambda}^{\infty}\frac{\mathrm{d}l}{l^{2}}J_{1}\left(k\left|\mathbf{\mathbf{v}}-\mathbf{\mathbf{v}'}\right|\right)J_{1}\left(l\left|\mathbf{\mathbf{v}}-\mathbf{\mathbf{v}'}\right|\right)\;,\\
 & =-\alpha_{s}^{2}\mu_{A}^{2}\left|\mathbf{\mathbf{v}}-\mathbf{\mathbf{v}'}\right|^{2}\left(1-2\gamma_{E}+\ln4+\ln\frac{1}{\left|\mathbf{\mathbf{v}}-\mathbf{\mathbf{v}'}\right|^{2}\Lambda^{2}}\right)^{2}\;,
\end{aligned}
\end{equation}
and
\begin{equation}
\begin{aligned} & \left.\left(\frac{2k_{t}^{i}k_{t}^{j}}{k_{t}^{2}}-\delta^{ij}\right)\frac{\partial}{\partial x^{i}}\Gamma\left(\mathbf{x}-\mathbf{v}'\right)\frac{\partial}{\partial y^{j}}\Gamma\left(\mathbf{v}-\mathbf{y}\right)\right|_{\mathbf{x}=\mathbf{v},\mathbf{y}=\mathbf{v}'}\\
 & =\left(2\mu_{A}g_{s}^{2}\right)^{2}\int\frac{\mathrm{d}^{2}\mathbf{q}}{\left(2\pi\right)^{2}}\int\frac{\mathrm{d}^{2}\mathbf{l}}{\left(2\pi\right)^{2}}\frac{1}{q^{4}l^{4}}\left(\frac{2\left(\mathbf{k}_{t}\cdot\mathbf{q}\right)\left(\mathbf{k}_{t}\cdot\mathbf{l}\right)}{k_{t}^{2}}-\mathbf{q}\cdot\mathbf{l}\right)e^{i\left(\mathbf{q}+\mathbf{l}\right)\cdot\left(\mathbf{\mathbf{v}}-\mathbf{\mathbf{v}'}\right)}\;,\\
 & =\frac{\left(2\mu_{A}g_{s}^{2}\right)^{2}}{\left(2\pi\right)^{4}}\int\frac{\mathrm{d}q\mathrm{d}\theta}{q^{2}}\int\frac{\mathrm{d}l\mathrm{d}\phi}{l^{2}}\left(2\cos\left(\theta+\alpha\right)\cos\left(\phi+\alpha\right)-\cos\left(\theta-\phi\right)\right)e^{iq\left|\mathbf{\mathbf{v}}-\mathbf{\mathbf{v}'}\right|\cos\left(\theta\right)}e^{il\left|\mathbf{\mathbf{v}}-\mathbf{\mathbf{v}'}\right|\mathbf{\cos\left(\phi\right)}}\;,\\
 & =-\frac{\left(2\mu_{A}g_{s}^{2}\right)^{2}}{\left(2\pi\right)^{2}}\int_{\Lambda}^{\infty}\frac{\mathrm{d}q}{q^{2}}\int_{\Lambda}^{\infty}\frac{\mathrm{d}l}{l^{2}}J_{1}\left(q\left|\mathbf{\mathbf{v}}-\mathbf{\mathbf{v}'}\right|\right)J_{1}\left(l\left|\mathbf{\mathbf{v}}-\mathbf{\mathbf{v}'}\right|\right)\cos\left(2\alpha\right)\;,\\
 & =\alpha_{s}^{2}\mu_{A}^{2}\left|\mathbf{\mathbf{v}}-\mathbf{\mathbf{v}'}\right|^{2}\cos\left(2\alpha\right)\left(1-2\gamma_{E}+\ln4+\ln\frac{1}{\left|\mathbf{\mathbf{v}}-\mathbf{\mathbf{v}'}\right|^{2}\Lambda^{2}}\right)^{2},
\end{aligned}
\end{equation}
from which, in combination with Eqs. \eqref{eq:I1} and \eqref{eq:I2}, one finds:
\begin{eqnarray}
\mathcal{F}_{gg}^{(1)}\left(x_{2},k_{t}\right) & = & \frac{S_{\perp}}{\alpha_{s}}\frac{C_{F}}{N_{c}^{2}}\frac{1}{32\pi^{3}}\int\mathrm{d}r\frac{J_{0}\left(k_{t}r\right)}{r}e^{-\frac{N_{c}}{2}\Gamma\left(r\right)}\nonumber \\&  & \Biggl[64\left(e^{\frac{N_{c}}{2}\Gamma\left(r\right)}-1\right)
  -\alpha_{s}^{2}N_{c}^{4}\mu_{A}^{2}r^{4}\left(1-2\gamma_{E}+\ln4+\ln\frac{1}{r^{2}\Lambda^{2}}\right)^{2}+8\alpha_{s}N_{c}\left(N_{c}^{2}-2\right)\mu_{A}r^{2}\ln\frac{1}{r^{2}\Lambda^{2}}\Biggr)\Biggr]\;\label{eq:F1MV}\\
\mathcal{H}_{gg}^{(1)}(x_{2},k_{t}) & = & \frac{S_{\perp}}{\alpha_{s}}\frac{C_{F}}{N_{c}^{2}}\frac{1}{32\pi^{3}}\int\mathrm{d}r\frac{J_{2}\left(k_{t}r\right)}{r}e^{-\frac{N_{c}}{2}\Gamma\left(r\right)}\nonumber \\  &  & \Biggl[\frac{64}{\ln\frac{1}{r^{2}\Lambda^{2}}}\left(e^{\frac{N_{c}}{2}\Gamma\left(r\right)}-1\right)
 +\alpha_{s}^{2}N_{c}^{4}\mu_{A}^{2}r^{4}\left(1-2\gamma_{E}+\ln4+\ln\frac{1}{r^{2}\Lambda^{2}}\right)^{2}+8\alpha_{s}N_{c}\left(N_{c}^{2}-2\right)\mu_{A}r^{2}\Biggr]\;,\\\label{eq:H1MV}
\mathcal{F}_{gg}^{\left(2\right)}\left(x_{2},k_{t}\right) & = & \frac{S_{\perp}}{\alpha_{s}}\frac{C_{F}}{N_{c}^{2}}\frac{1}{32\pi^{3}}\int\mathrm{d}r\frac{J_{0}\left(k_{t}r\right)}{r}e^{-\frac{N_{c}}{2}\Gamma\left(r\right)}\nonumber \\  &  &\Biggl[64\left(e^{\frac{N_{c}}{2}\Gamma\left(r\right)}-1\right)
 +\alpha_{s}^{2}N_{c}^{4}\mu_{A}^{2}r^{4}\left(1-2\gamma_{E}+\ln4+\ln\frac{1}{r^{2}\Lambda^{2}}\right)^{2}-16\alpha_{s}N_{c}\mu_{A}r^{2}\ln\frac{1}{r^{2}\Lambda^{2}}\Biggr]\;,\label{eq:F2MV}\\
\mathcal{H}_{gg}^{\left(2\right)}\left(x_{2},k_{t}\right) & = & \frac{S_{\perp}}{\alpha_{s}}\frac{C_{F}}{N_{c}^{2}}\frac{1}{32\pi^{3}}\int\mathrm{d}r\frac{J_{2}\left(k_{t}r\right)}{r}e^{-\frac{N_{c}}{2}\Gamma\left(r\right)}\nonumber \\
 &  & \Biggl[\frac{64}{\ln\frac{1}{r^{2}\Lambda^{2}}}\left(e^{\frac{N_{c}}{2}\Gamma\left(r\right)}-1\right)-\alpha_{s}^{2}N_{c}^{4}\mu_{A}^{2}r^{4}\left(1-2\gamma_{E}+\ln4+\ln\frac{1}{r^{2}\Lambda^{2}}\right)^{2}-16\alpha_{s}N_{c}\mu_{A}r^{2}\Biggr]\;.\label{eq:H2MV}
\end{eqnarray}

\subsection{Comparison with the literature}
In order to recover the results found in \cite{Akcakaya:2012si} from our cross section \eqref{eq:finalCGCcrosssection}, one needs to introduce the following 
auxiliary TMDs $xG_{q\bar{q}}\left(x_{2},k_{t}\right)$
and $xH_{q\bar{q}}\left(x_{2},k_{t}\right)$, defined as:
\begin{equation}
\begin{aligned}xG_{q\bar{q}}\left(x_{2},k_{t}\right) & \equiv\frac{S_{\perp}N_{c}}{2\pi^{2}\alpha_{s}}\int\frac{\mathrm{d}^{2}\mathbf{r}}{\left(2\pi\right)^{2}}Q_{s}^{2}\left(r^{2}\right)e^{-i\mathbf{k}_{t}\cdot\mathbf{r}}e^{-\frac{N_{c}}{2}\Gamma\left(\mathbf{r}\right)}\;,\\
xH_{q\bar{q}}\left(x_{2},k_{t}\right) & \equiv\frac{N_{c}^{2}-1}{8\pi^{3}}S_{\perp}\mu_{A}\int\mathrm{d}r\,rJ_{2}\left(k_{t}r\right)e^{-\frac{N_{c}}{2}\Gamma\left(\mathbf{r}\right)}\;.
\end{aligned}
\label{eq:xGqq}
\end{equation}
From \eqref{eq:F1MV}-\eqref{eq:H2MV}, it is straightforward to derive
\begin{equation}
\begin{aligned}\mathcal{F}_{gg}^{\left(1\right)}\left(x_{2},k_{t}\right)+\mathcal{F}_{gg}^{\left(2\right)}\left(x_{2},k_{t}\right) & =\frac{4}{N_{c}^{2}}\mathcal{F}_{gg}^{\left(3\right)}\left(x_{2},k_{t}\right)+\left(1-\frac{4}{N_{c}^{2}}\right)xG_{q\bar{q}}\left(x_{2},k_{t}\right)\;,\\
\mathcal{H}_{gg}^{\left(1\right)}\left(x_{2},k_{t}\right)+\mathcal{H}_{gg}^{\left(2\right)}\left(x_{2},k_{t}\right) & =\frac{4}{N_{c}^{2}}\mathcal{H}_{gg}^{\left(3\right)}\left(x_{2},k_{t}\right)+\left(1-\frac{4}{N_{c}^{2}}\right)xH_{q\bar{q}}\left(x_{2},k_{t}\right)\ ,
\end{aligned}
\label{eq:plus}
\end{equation}
and to write the unpolarized and linearly polarized part of the cross section \eqref{eq:finalCGCcrosssection} in the following way (to leading order in $m^2/P_t^2$): 
\begin{equation}
\begin{aligned} & \left.\frac{\mathrm{d}\sigma^{pA\rightarrow Q\bar{Q}X}}{\mathrm{d}\mathcal{P}.\mathcal{S}.}\right|_{unp.}=\frac{\alpha_{s}^{2}}{4C_{F}}\frac{z(1-z)}{P_t^{4}}x_{1}g\left(x_{1},\mu^{2}\right)P_{qg}(z)\\
 & \times\Biggl\{ (1-2z)^2 \mathcal{F}_{ADP}\left(x_2,k_{t}\right)+\left(1-\frac{4}{N_{c}^{2}}\right)xG_{q\bar{q}}\left(x_2,k_{t}\right)+\frac{2}{N_{c}^{2}}\mathcal{F}_{gg}^{\left(3\right)}\left(x_2,k_{t}\right)\Biggr\},\\
 \left.\frac{\mathrm{d}\sigma^{pA\rightarrow Q\bar{Q}X}}{\mathrm{d}\mathcal{P}.\mathcal{S}.}\right|_{pol.} & =\frac{\alpha_{s}^{2}}{2C_{F}}\frac{z^2(1-z)^2}{P^{4}_t}x_{1}g\left(x_{1},\mu^{2}\right)\frac{m^{2}}{P_{t}^{2}}\cos\left(2\phi\right)\\
 & \times\Biggl\{(1-2z)^2\mathcal{F}_{ADP}\left(x_2,k_{t}\right)+\left(1-\frac{4}{N_{c}^{2}}\right)xH_{q\bar{q}}\left(x_2,k_{t}\right)+\frac{2}{N_{c}^{2}}\mathcal{H}_{gg}^{\left(3\right)}\left(x_2,k_{t}\right)\Biggr\}.
\end{aligned}
\end{equation}
These are the expressions derived in \cite{Akcakaya:2012si}. They are valid in the MV model (without the need to neglect the $\ln(r)$ dependence of the saturation scale) only, as they make use of \eqref{eq:plus}, but they are not simpler than the general form we obtained in Section II (the same number of TMDs are involved).

\section{JIMWLK evolution of the linearly-polarized gluon TMDs}
\label{sec:lattice}

\subsection{Numerical implementation}

The JIMWLK evolution equation in rapidity, \D{y=\ln\lp{x_0/x}}, can be
solved at fixed coupling in the small-$x$ regime on a two-dimensional
lattice with a Langevin diffusion process of $SU(3)$ matrix variables
\cite{Weigert:2000gi}. The matrix degrees of freedom represent
partonic Wilson lines along a light-cone direction and the lattice
discretizes transverse space. We use the numerical code developed for
the calculation of unpolarized gluon TMDs in
\cite{Marquet:2016cgx}. This code is based on algorithms described by
Rummukainen and Weigert \cite{Rummukainen:2003ns} and Lappi
\cite{Lappi:2007ku,Lappi:2012vw}. We choose to generate the initial
$SU(3)$ configurations at $y=0$ in the McLerran-Venugopalan (MV) model
wherein analytical calculations of the gluon TMDs have been
performed in the previous section.

The gluon distributions listed in Eqs.~\eqref{eq:F1}-\eqref{eq:H3} are two-point functions defined as products of various
traces, which must be evaluated component-wise with respect to the
spatial and color indices, in order to express them as scalar
convolution products which can be calculated efficiently using a
discrete fast Fourier transform algorithm. The continuum derivative
$\d_\alpha$ can be replaced either by a discrete forward or a central
difference operator $\nabla_\alpha$. The most convenient expressions
for a numerical implementation on a square lattice of size $L\times L$
are the following formulas for the unpolarized gluon distributions:
\begin{equation}
\label{eq:LF123}
\begin{aligned}
\FC_{gg}^{(1)}\left(x,k_t\right)
& =\f{1}{2\pi^3g_{s}^{2}}\f{1}{N_{c}}
\sum_{\alpha=1}^{2}\sum_{i,j,k,l=1}^{N_{c}}
\Biggl\langle\biggl|
\sum_{\b{v}} e^{i\b{k}_{t}\cdot\b{v}}
U_{ij}^{\dagger}(\b{v})\nabla_{\alpha}U_{kl}(\b{v})
\biggr|^{2}\Biggr\rangle_{x}\,,\\
\FC_{gg}^{(2)}\left(x,k_t\right)
& =-\f{1}{2\pi^3g_{s}^{2}}\f{1}{N_{c}}
\sum_{\alpha=1}^{2}\sum_{i,j,k,l=1}^{N_{c}}
\mathrm{Re}\Biggl\langle
\biggl(\sum_{\b{v}} e^{-i\b{k}_{t}\cdot\b{v}}
U_{ij}^{\dagger}(\b{v})\nabla_{\alpha}U_{kl}(\b{v})\biggr)\\
&\times\biggl(\sum_{\b{v}} e^{i\b{k}_{t}\cdot\b{v}}
U_{lk}^{\dagger}(\b{v})\nabla_{\alpha}U_{ji}(\b{v})\biggr)
\Biggr\rangle_{x}\,,\\
\FC_{gg}^{(3)}\left(x,k_t\right)
& =\f{1}{2\pi^3g_{s}^{2}}\sum_{\alpha=1}^{2}
\sum_{i,j}^{N_{c}}\Biggl\langle\biggl|
\sum_{\b{v}} e^{-i\b{k}_{t}\cdot\b{v}}
\left(U^{\dagger}(\b{v})\nabla_{\alpha}U(\b{v})\right)_{ij}
\biggr|^{2}\Biggr\rangle_{x}\,,
\end{aligned}
\end{equation}
and for the linearly polarized distributions:
\begin{equation}
\label{eq:LH123}
\begin{aligned}
\HC_{gg}^{(1)}\left(x,k_t\right) 
& =\f{1}{\pi^3g_{s}^{2}}\f{1}{N_{c}}
\sum_{i,j,k,l=1}^{N_{c}}
\Biggl\langle\biggr|\sum_{\alpha=1}^2
\f{\w{k}_{t}^{\alpha}}{\w{k}_{t}}
\sum_{\b{v}} e^{i\b{k}_{t}\cdot\b{v}}
U_{ij}^{\dagger}(\b{v})\nabla_{\alpha}U_{kl}(\b{v})
\biggr|^{2}\Biggr\rangle_{x}
-\FC_{gg}^{(1)}\left(x,k_t\right)\,,\\
\HC_{gg}^{(2)}\left(x,k_t\right) 
& =-\f{1}{\pi^3g_{s}^{2}}\f{1}{N_{c}}
\sum_{i,j,k,l=1}^{N_{c}}
\mathrm{Re}\Biggl\langle
\biggl(\sum_{\alpha=1}^2
\f{\w{k}_{t}^{\alpha}}{\w{k}_{t}}
\sum_{\b{v}} e^{-i\b{k}_{t}\cdot\b{v}}
U^{\dagger}_{ij}(\b{v})\nabla_{\alpha}U_{kl}(\b{v})\biggr)\\
& \times\biggl(\sum_{\alpha=1}^2
\f{\w{k}_{t}^{\alpha}}{\w{k}_{t}}
\sum_{\b{v}} e^{i\b{k}_{t}\cdot\b{v}}
U_{lk}^{\dagger}(\b{v})\nabla_{\alpha}U_{ji}(\b{v})\biggr)
\Biggr\rangle_{x}
-\FC_{gg}^{(2)}\left(x,k_t\right)\,,\\
\HC_{gg}^{(3)}\left(x,k_t\right) 
& =\f{1}{\pi^3g_{s}^{2}}\sum_{i,j=1}^{N_{c}}
\Biggl\langle\biggr|\sum_{\alpha=1}^2
\f{\w{k}_{t}^{\alpha}}{\w{k}_{t}}
\sum_{\b{v}} e^{-i\b{k}_{t}\cdot\b{v}}
\left(U^{\dagger}_{\b{v}}\nabla_{\alpha}U_{\b{v}}\right)_{ij}
\biggr|^{2}\Biggr\rangle_{x}
-\FC_{gg}^{(3)}\left(x,k_t\right)\,,
\end{aligned}
\end{equation}
where $k_t^\alpha$ is the momentum in the lattice Brillouin zone and $\w{k}_t^\alpha$ is either the forward lattice momentum
$\left(\h{k}^\alpha_t=2\sin\f{k^\alpha_t}{2}\right)$ or
central lattice momentum $\left(\h{\h{k^\alpha_t}}=\sin
  k^\alpha_t\right)$, in accordance with the definition of the
difference operator $\nabla^\alpha$.

All these lattice gluon distributions have the correct continuum limit
and are thus expressed in terms of $2N_c^4$ complex two-dimensional
discrete Fourier transforms.

\subsection{Adjoint sum rules}

With the above definitions, the continuum sum rules
\eqref{eq:F1minusF2} and \eqref{eq:H1minusH2} remain true on the
lattice. The dipole correlator in the adjoint representation is
defined as
\begin{align}
  \label{eq:CA}
  \begin{split}
  S_{gg}^{(2)}(\b{v}-\b{v}';x) &= \frac{1}{N_c^2-1}
  \vev{\left| \text{Tr}\, U^{\dagger}_{\b{v}}U_{\b{v}'}\right|^2-1}_x  \\
  &= \frac{1}{N_c^2-1} \sum_{i,j,k,l=1}^{N_c} \vev{\lp{U_{ij}(\b{v})U_{kl}^{*}(\b{v})}
    \lp{U_{ij}(\b{v}')U^{*}_{kl}(\b{v}')}^{*}- 1 }_x \,.
  \end{split}
\end{align}
Taking the second-order lattice cross-derivative with respect to
$v_\alpha$ and $v'_\alpha$ of the R.H.S. of \eqref{eq:CA} we recognize
in the various terms, after some interchange $(ij)\leftrightarrow(kl)$
of dummy color indices and/or dummy spatial indices
$\b{v}\leftrightarrow\b{v}'$, the contributions to
$\FC^{(1)}(x,k_{t})$ and $\FC^{(2)}(x,k_{t})$ in
\eqref{eq:LF123}. Hence we have a lattice sum rule which holds in fact
configuration by configuration on a finite periodic lattice where the
adjoint correlator $S_{gg}^{(2)}(\b{v}-\b{v}';x) $ is translation-invariant:
\begin{align}
\label{eq:CSR}
\f{C_F}{2\pi^3g^2_s}\sum_{\alpha=1}^2\sum_{\b{v},\b{v}'}
e^{i\b{k}_{t}\cdot(\b{v}-\b{v'})}
\nabla_{v^{\alpha}}\nabla_{v'^{\alpha}} S_{gg}^{(2)}(\b{v}-\b{v}';x)  &=
\FC^{(1)}(x,k_t) - \FC^{(2)}(x,k_t)\,.
\end{align}
This sum rule remains valid whether one defines the lattice derivative
as a forward ($\w{k}_t^2\equiv\h{k}_t^2$) or central
difference operator ($\w{k}_t^2\equiv\h{\h{k}}_t^2$). The Fourier
transformations of the lattice derivatives in the R.H.S of
eqs.\,\eqref{eq:LH123} and L.H.S of \eqref{eq:CSR} yield the two
lattice adjoint sum rules:
\begin{align}
  \label{eq:DSR}
  \begin{split}
  \f{C_F L^2}{2\pi^3g^2_s}\w{k}_t^2\h{S}_{gg}^{(2)}(x,k_t)
  &= \FC^{(1)}(x,k_t) - \FC^{(2)}(x,k_t)\,, \\
  &= \HC^{(1)}(x,k_t) - \HC^{(2)}(x,k_t)\,,
  \end{split}
\end{align}
\begin{equation}
  \h{k}_t^2 = 4\sum_{\alpha=1}^2
  \sin^2\f{k^\alpha_t}{2}\,,\quad
  \h{\h{k}}_t^2 = \sum_{\alpha=1}^2\sin^2 k^\alpha_t\,,\quad
  \h{S}_{gg}^{(2)}(x,k_t) = \sum_{\b{z}}e^{-i\b{k}_{t}\cdot\b{z}}S_{gg}^{(2)}(\b{z};x)\,.
\end{equation}
The lattice adjoint sum rules hold true configuration by configuration
only if one uses the same definition of the lattice derivative in both
sides of \eqref{eq:DSR}. Since it requires only one additional discrete Fourier transform from a single Langevin simulation, using a different definition provides a very
economical way to measure the growth of lattice (and rapidity) discretization errors during the JIMWLK evolution.

\subsection{JIMWLK evolution}

All numerical measurements of the gluon distributions studied in this
work have been performed on the lattice size $L=1024$. A statistical
sample of 50 independent trajectories in rapidity has been generated
with initial $SU(3)$ configurations distributed so that the
correlation length of the dipole correlator in the fundamental
representation be in lattice units:
\begin{align}
  \label{eq:Rs}
  \vev{R_s}_{x_0}= 65.8 \pm 0.3\,.
\end{align}
This correlation length is defined following the standard Gaussian-like convention:
\begin{align}
  \label{eq:Dipole} 
  S_{q\bar{q}}(R_s;x) =e^{-1/2} \,.
\end{align}
It is indeed unreliable to study lattice momenta beyond $\f{\pi}{4}$
in the Brillouin zone of a periodic lattice and a safe upper bound for
the initial correlation length is
$\displaystyle{R_s\lesssim\frac{L}{16}}$ (the correlation length then decreases with evolution). We treat the issue of those
lattice artifacts in momentum space which are due to the breaking of $O(2)$ rotational invariance by a square lattice
 exactly as explained in \cite{Marquet:2016cgx}. In all subsequent figures, we choose to
keep only the (integer) momenta $k=(k_1,k_2)$ with
\begin{align}
  \label{eq:cut}
  |k_1-k_2| \leq 5\,.
\end{align}

All averages or data points displayed in the plots have error bars
which have been determined from the JIMWLK evolution of the random
sample with a Langevin step $\displaystyle{\delta s =
  \frac{\alpha_s}{\pi^2}\delta y = 10^{-4}}$. To interpret the results
it is convenient to fix physical units. If we assume a starting
$x$ of $x_{0}=10^{-2}$, with associated saturation scale
$Q_{s}^{2}\left(x_{0}\right)=0.2\,\mathrm{GeV}^{2}$, we can restore
the lattice spacing $a$ from \eqref{eq:Rs} and find:
\begin{equation}
a=\frac{\sqrt{2}}{66\times\sqrt{0.2}}\mathrm{GeV}^{-1}\simeq0.05\,
\mathrm{GeV}^{-1}.
\end{equation}
Choosing $\alpha_{s}=0.15$, a value of
$\left(\alpha_{s}/\pi^{2}\right)y=0.1$ would correspond to
$x\simeq 1.4\ 10^{-5}$.

\begin{figure}
  \begin{center}
    \includegraphics[width=0.75\textwidth]{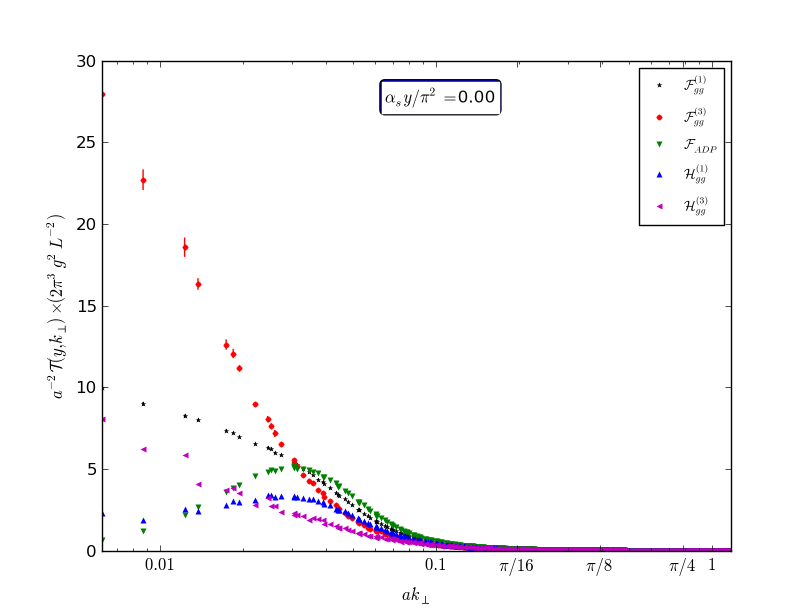}
  \end{center}
  \caption{Momentum dependence of gluon TMDs in the initial MV
    configurations. The linear vertical scale is rescaled by the
    factor $2\pi^3g^2L^{-2}$. The logarithmic momentum scale is in inverse
    lattice spacing units. $\mathcal{T}(y,k_t)$ is a shorthand
    for the labels of the gluon TMDs displayed in the figure.}
  \label{fig:TMDs_0}
\end{figure}

Fig.~\ref{fig:TMDs_0} displays the initial gluon TMDs calculated on
the lattice in the MV model at $y=0$. As already observed in \cite{Marquet:2016cgx} for the unpolarized gluon TMDs
$\FC^{(1)}_{gg}$ and $\FC^{(3)}_{gg}$, the linearly polarized gluon
TMDs $\HC^{(1)}_{gg}$ and $\HC^{(3)}_{gg}$, as well as the adjoint
dipole correlator $\FC_{ADP}$, have also the expected universal
$1/k^2_t$ behavior at large $k_t$ \cite{vanHameren:2016ftb}.

\begin{figure}
  \begin{center}
    \includegraphics[width=0.75\textwidth]{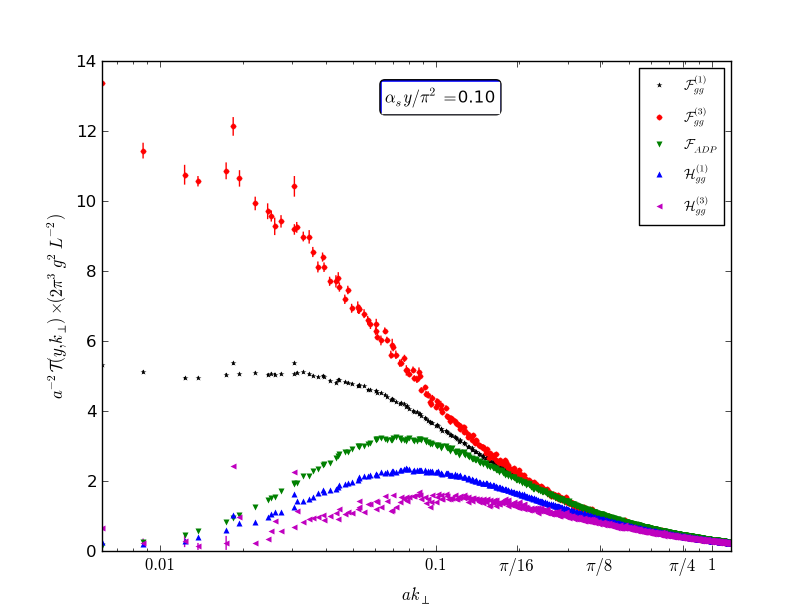}
    \includegraphics[width=0.75\textwidth]{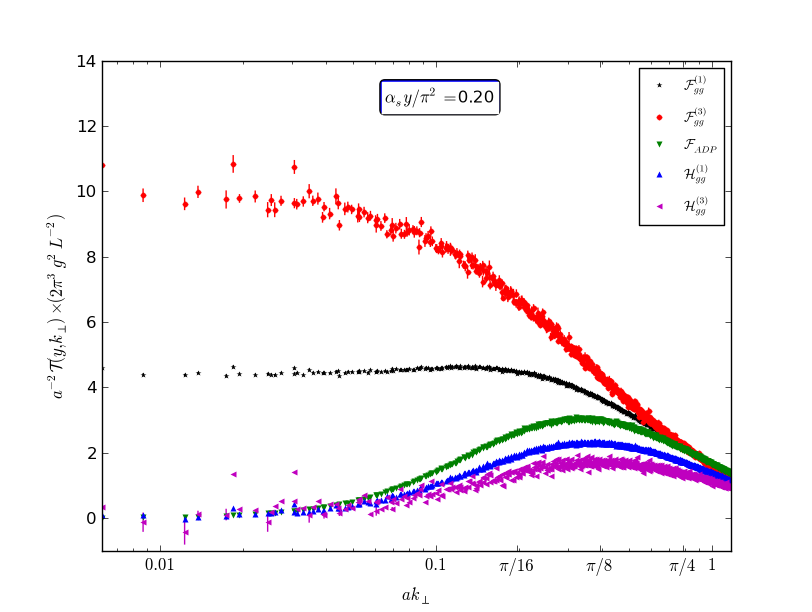}
  \end{center}
  \caption{Momentum dependence of gluon TMDs near the lower bound
    (top) and upper bound (bottom) of the geometric scaling
    window. The logarithmic momentum scale is in inverse lattice
    spacing units. $\mathcal{T}(y,k_t)$ is a shorthand
    for the labels of the gluon TMDs displayed in the figure.}
  \label{fig:TMDs_evolved}
\end{figure}

The high-$k_{t}$ behavior during the JIMWLK evolution is best
exhibited by looking at the top plot of Fig.~\ref{fig:TMDs_evolved},
which shows the gluon TMDs for $\alpha_s y/\pi^2=0.1$, after enough
evolution to have reached the geometric scaling regime, but not too
much so that the high-$k_{t}$ tails of the gluon distributions
stays within the accessible momentum range on the lattice. Our results
for $\mathcal{F}_{gg}^{\left(1\right)}$ and
$\mathcal{F}_{gg}^{\left(3\right)}$ match
indeed with \cite{Marquet:2016cgx}. Furthermore, we reproduce,
at least qualitatively, the numerical results for
$\mathcal{F}_{gg}^{\left(3\right)}$ and
$\mathcal{H}_{gg}^{\left(3\right)}$ in
\cite{Dumitru:2015gaa}. The important observation to be made is that the data confirm the observations made earlier in
\cite{Marquet:2016cgx}, that in the limit of
large $k_{t}$, the high-energy or $k_{t}$-factorization regime is
recovered, in which all gluon TMDs converge to a common unintegrated
PDF. The only exceptions are the distributions
$\mathcal{F}_{gg}^{\left(2\right)}$ and
$\mathcal{H}_{gg}^{\left(2\right)}$, which
vanish very fast. They are not shown on the figures, instead 
$\FC_{ADP}(=\FC^{(1)}_{gg}-\FC^{(2)}_{gg}=\HC^{(1)}_{gg}-\HC^{(2)}_{gg})$ is plotted.

We also show in the bottom of Fig.~\ref{fig:TMDs_evolved}, the gluon
TMDs after further evolution at $\alpha_sy/\pi^2=0.2$, where the
high $k_t$ has disappeared from the accessible momentum range of
our analysis. On the other hand, this allows us to probe the
saturation regime at low $k_t$, where the various gluon TMDs are
very different from each other, and where the process dependence of
TMDs is most relevant and cannot be ignored.

Furthermore, when evolving towards smaller values of $x$, the gluon TMDs shift towards larger values of $k_{t}$. This is to be
expected from the fact that the distributions follow the saturation
scale, which grows when $x$
decreases. The different values of $\left(\alpha_{s}/\pi^{2}\right)y$, for which we plot the numerical
results, are listed in Table \ref{tab:SvsX}, along with the corresponding value of $x$, as well as the approximate value of
the saturation scale $Q_{sg}$ directly measured from the maximum of the adjoint dipole distribution (this method of extraction implies
small differences with the values of $\sqrt{N_c/C_F}\ Q_s\!=\!1.5\ Q_s$ obtained with the definition \eqref{eq:Dipole}).
\begin{table}[h]
\begin{centering}
\begin{tabular}{|c|c|c|}
\hline 
$\left(\alpha_{s}/\pi^{2}\right)y$ & $x$ & $Q_{sg}$\tabularnewline
\hline 
\hline 
$0$ & $x_{0}=10^{-2}$ & $0.6\,\mathrm{\mathrm{GeV}}$\tabularnewline
\hline 
$0.1$ & $1.4\cdot 10^{-5}$ & $1.5\,\mathrm{\mathrm{GeV}}$\tabularnewline
\hline 
$0.2$ & $2\cdot10^{-8}$ & $7\,\mathrm{\mathrm{GeV}}$\tabularnewline
\hline 
\end{tabular}
\par\end{centering}
\caption{\label{tab:SvsX}The values of $\left(\alpha_{s}/\pi^{2}\right)y$
versus $x$, and the corresponding value of the saturation
scale, calculated from the maximum of $\FC_{ADP}$.}
\end{table}

Regarding the information on the gluon polarization, $\mathcal{H}_{gg}^{\left(1\right)}$ and
$\mathcal{H}_{gg}^{\left(2\right)}$ are small to begin with, but the magnitude of $\mathcal{H}_{gg}^{\left(3\right)}$ is comparable to that of $\mathcal{F}_{gg}^{\left(1\right)}$ (and $\mathcal{F}_{gg}^{\left(2\right)}$) for small values of $k_t$. There, $\mathcal{F}_{gg}^{\left(1\right)}$
is equal to $\mathcal{F}_{gg}^{\left(2\right)}$ and $\mathcal{H}_{gg}^{\left(1\right)}$ is equal
to $\mathcal{H}_{gg}^{\left(2\right)}$; this is a consequence of the adjoint sum rules \eqref{eq:DSR} in combination with the fact that the adjoint dipole correlator
$\FC_{ADP}$ vanishes as $k_t^2$ for $k_t=0$. The linearly-polarized gluons get suppressed after the first steps in the evolution, however, they are not completely washed out. Indeed, the distributions of linearly-polarized gluons all remain non-zero for momenta of the order of the saturation scale.

\section{Conclusions}

In this paper, we used the CGC framework to compute the cross section for the forward production of a heavy quark-antiquark pair in proton-nucleus collisions. In the correlation limit, in which the outgoing quarks are almost back-to-back in the transverse plane, our result could be cast into a TMD factorization formula, involving six different gluon TMDs. Three of these TMDs are unpolarized, and also appear in the cross section for forward dijet production. They are each accompanied by a partner which couples via the quark mass, and which corresponds to the linearly-polarized gluons inside the unpolarized nucleus. We have obtained analytical expressions for each of the TMDs in the MV model. Furthermore, the gluon TMDs were numerically evolved in rapidity using the nonlinear JIMWLK evolution equation.

Our results indicate that the various distributions of linearly-polarized gluons always remain non-zero for values of the gluon transverse momentum of the order of the saturation scale. This observation provides us with a novel way to test parton saturation at the LHC, and to extract the poorly known linearly-polarized gluon distributions. We note that the LHCb detector would be particularly well suited to perform such a measurement of heavy mesons in the forward region, although a detailed feasibility study remains to be done. Indeed, the linearly-polarized gluons impact the production cross section via an angular modulation whose magnitude we plan to better quantify.

In appendix \ref{sec:DIS}, we give an outline of the derivation for a similar but simpler process, $\gamma^* A\rightarrow Q\bar{Q}X$, in which only two of the gluon TMDs appear: the Weizs\"acker-Willams distribution $\mathcal{F}_{gg}^{\left(3\right)}\left(x,k_{t}\right)$ and its polarized partner $\mathcal{H}_{gg}^{\left(3\right)}\left(x,k_{t}\right)$. Interestingly, the dependence on $\mathcal{H}_{gg}^{\left(3\right)}\left(x,k_{t}\right)$ via the azimuthal angle between $P_t$ and $k_t$, couples not only to the quark mass but also to the virtuality of the photon. This provides alternatives to the $pA\to Q\bar{Q}X$ process -- namely dijets at an Electron-Ion Collider \cite{Boer:2016fqd}, or heavy $Q\bar Q$ pair production in ultra-peripheral collisions of heavy ions -- using processes whose theoretical formulation involves less gluon TMDs, but which may be experimentally more challenging or more distant in the future.

Finally, let us stress again that the focus of this work was on the implementation of the small-$x$ JIMWLK evolution, and that we did not discuss the scale evolution. That aspect was recently studied in the simpler context of the $pA\to \gamma^*qX$ process \cite{Boer:2017xpy}, and it was found that the scale evolution leads to a Sudakov suppression of the angular modulation induced by the linear polarization of gluons. However, in that process only the fundamental-dipole gluon TMD is involved, which is a peculiar TMD since, as we already pointed out, the unpolarized and linearly-polarized distributions are identical at small-$x$. We leave it for future work to estimate the effect of the scale evolution on the TMDs displayed in Fig.~\ref{fig:TMDs_evolved}, involved in the $pA\to Q\bar{Q}X$ process. We also note that, at next-to-leading order, additional gluon TMDs appear \cite{Benic:2017znu} in the $pA\to \gamma^*qX$ process, related to those discussed in this work.

\section*{Acknowledgments}

The work of CM was supported in part by the Agence Nationale de la Recherche under the project ANR-16-CE31-0019-02.
The research of PT has been partially funded by NCN grant DEC-2013/10/E/ST2/00656 and by the FWO-PAS grant.

\section*{Appendices}
\appendix

\section{\label{sec:DIS}Massive forward heavy-quark pair production in deep-inelastic scattering}

The cross section for dihadron production in deep-inelastic scattering
reads \cite{Dominguez:2011wm}:
\begin{equation}
\begin{aligned}\frac{\mathrm{d}\sigma^{\gamma^* A\rightarrow Q\bar{Q}X}}{\mathrm{d}y_{1}\mathrm{d}y_{2}\mathrm{d}^{2}p_{1t}\mathrm{d}^{2}p_{2t}} & =N_{c}\alpha_{em}e_{q}^{2}z\left(1-z\right)\delta\left(1-\frac{p_{1}^{+}+p_{2}^{+}}{p^+}\right)\int\frac{\mathrm{d}^{2}\mathbf{u}}{\left(2\pi\right)^{2}}\frac{\mathrm{d}^{2}\mathbf{u}'}{\left(2\pi\right)^{2}}\frac{\mathrm{d}^{2}\mathbf{v}}{\left(2\pi\right)^{2}}\frac{\mathrm{d}^{2}\mathbf{v}'}{\left(2\pi\right)^{2}}\\
 & \times e^{-i\mathbf{k}_{t}\cdot\left(\mathbf{v}-\mathbf{v}'\right)}e^{-i\mathbf{P}_{t}\cdot\left(\mathbf{u}-\mathbf{u}'\right)}p^{+}\sum_{\lambda\alpha\beta}\psi_{\alpha\beta}^{L,T\lambda}\left(\mathbf{u}\right)\psi_{\alpha\beta}^{L,T\lambda*}\left(\mathbf{u}'\right)\\
 & \times\left[1+\langle Q\left(\mathbf{x},\mathbf{x}',\mathbf{b}',\mathbf{b}\right) \rangle_{x_2}-\langle D\left(\mathbf{x},\mathbf{b}\right) \rangle_{x_2}-\langle D\left(\mathbf{x}',\mathbf{b}'\right) \rangle_{x_2} \right]\;.
\end{aligned}
\label{eq:startDIS}
\end{equation}
The overlap of the wave functions of the longitudinally and transversally
polarized photon is given by, respectively:
\begin{equation}
\begin{aligned}p^{+}\sum_{\alpha\beta}\psi_{\alpha\beta}^{L}\left(\mathbf{u}\right)\psi_{\alpha\beta}^{L*}\left(\mathbf{u}'\right) & =16\pi^{2}Q^{2}z^{2}\left(1-z\right)^{2}\sum_{\alpha\beta}K_{0}\left(\epsilon_{f}u\right)K_{0}\left(\epsilon_{f}u'\right)\;,\end{aligned}
\end{equation}
and
\begin{equation}
\frac{p^{+}}{2}\sum_{\lambda=1,2}\sum_{\alpha\beta}\psi_{\alpha\beta}^{\lambda}\left(u\right)\psi_{\alpha\beta}^{\lambda*}\left(u'\right)=4\pi^{2}\left[\epsilon_{f}^{2}K_{1}\left(\epsilon_{f}u\right)K_{1}\left(\epsilon_{f}u'\right)\frac{\mathbf{u}\cdot\mathbf{u}'}{uu'}\left(z^{2}+\left(1-z\right)^{2}\right)+m^{2}K_{0}\left(\epsilon_{f}u\right)K_{0}\left(\epsilon_{f}u'\right)\right]\;,
\end{equation}
where:
\begin{equation}
\epsilon_{f}^{2}=m^{2}+z\left(1-z\right)Q^{2}\;.
\end{equation}

Following the same procedure as in section \ref{sec:correlationlimit}, taking the correlation limit, one obtains the following factorization formulae:
\begin{equation}
\begin{aligned}\frac{\mathrm{d}\sigma^{\gamma^{*}_L A\rightarrow Q\bar{Q}X}}{\mathrm{d}y_{1}\mathrm{d}y_{2}\mathrm{d}^{2}p_{1t}\mathrm{d}^{2}p_{2t}} & =8\alpha_{s}\alpha_{em}e_{q}^{2}Q^2 \delta\left(1-\frac{p_{1}^{+}+p_{2}^{+}}{p^{+}}\right)z^{3}\left(1-z\right)^{3}\frac{P_{t}^{2}}{\left(P_{t}^{2}+\epsilon_{f}^{2}\right)^{4}} \\& \times  \left(\mathcal{F}_{gg}^{\left(3\right)}\left(x_{2},k_{t}\right)+\cos\left(2\phi\right)\mathcal{H}_{gg}^{\left(3\right)}\left(x_{2},k_{t}\right)\right),\end{aligned}
\label{eq:longitudinalresultphi}
\end{equation}
and
\begin{equation}
\begin{aligned}\frac{\mathrm{d}\sigma^{\gamma^{*}_T A\rightarrow Q\bar{Q}X}}{\mathrm{d}y_{1}\mathrm{d}y_{2}\mathrm{d}^{2}p_{1t}\mathrm{d}^{2}p_{2t}} & =\alpha_{s}\alpha_{em}e_{q}^{2}\delta\left(1-\frac{p_{1}^{+}+p_{2}^{+}}{p^{+}}\right)z\left(1-z\right)\frac{1}{\left(P_{t}^{2}+\epsilon_{f}^{2}\right)^{4}}\\
 & \times\left\{ \Biggl[\left(P_{t}^{4}+\epsilon_{f}^{4}\right)\left(z^{2}+\left(1-z\right)^{2}\right)+2m^{2}P_{t}^{2}\Biggr]\mathcal{F}_{gg}^{\left(3\right)}\left(x_{2},k_{t}\right)\right.\\
 & +\left.\Biggl[-2\epsilon_{f}^{2}P_{t}^{2}\left(z^{2}+\left(1-z\right)^{2}\right)+2m^{2}P_{t}^{2}\Biggr]\cos\left(2\phi\right)\mathcal{H}_{gg}^{\left(3\right)}\left(x_{2},k_{t}\right)\right\} \;,
\end{aligned}
\label{eq:transverseresultphi}
\end{equation}
where we made use of the following integrals:
\begin{equation}
\int\frac{\mathrm{d}^{2}\mathbf{u}}{\left(2\pi\right)^{2}}\frac{\mathrm{d}^{2}\mathbf{u}'}{\left(2\pi\right)^{2}}e^{-i\mathbf{P}_{t}\cdot\left(\mathbf{u}-\mathbf{u}'\right)}u_{i}u'_{j}p^{+}\sum_{\alpha\beta}\psi_{\alpha\beta}^{L}\left(u\right)\psi_{\alpha\beta}^{L*}\left(u'\right)  =32Q^{2}z^{2}\left(1-z\right)^{2}\frac{P_{i}P_{j}}{\left(P_{t}^{2}+\epsilon_{f}^{2}\right)^{4}},\end{equation}
\begin{equation}
\begin{aligned} & \int\frac{\mathrm{d}^{2}\mathbf{u}}{\left(2\pi\right)^{2}}\frac{\mathrm{d}^{2}\mathbf{u}'}{\left(2\pi\right)^{2}}e^{-i\mathbf{P}_{t}\cdot\left(\mathbf{u}-\mathbf{u}'\right)}u_{i}u'_{j}p^{+}\sum_{\lambda=1,2}\sum_{\alpha\beta}\psi_{\alpha\beta}^{\lambda}\left(u\right)\psi_{\alpha\beta}^{\lambda*}\left(u'\right)\\
 & =4\left(\frac{\delta_{ij}}{\left(P_{t}^{2}+\epsilon_{f}^{2}\right)^{2}}-\frac{4\epsilon_{f}^{2}P_{i}P_{j}}{\left(P_{t}^{2}+\epsilon_{f}^{2}\right)^{4}}\right)\left(z^{2}+\left(1-z\right)^{2}\right)+\frac{16m^{2}P_{i}P_{j}}{\left(P_{t}^{2}+\epsilon_{f}^{2}\right)^{4}}\;.
\end{aligned}
\end{equation}
The cross sections \eqref{eq:longitudinalresultphi} and \eqref{eq:transverseresultphi} were first obtained in \cite{Metz:2011wb}. Recently, the next-to-leading power was also obtained in the massless limit \cite{Dumitru:2016jku}.

\section{\label{sec:operatordef}Transverse momentum dependent gluon distributions}

\begin{figure}[t]
\begin{center}
 \includegraphics[width=0.4\textwidth]{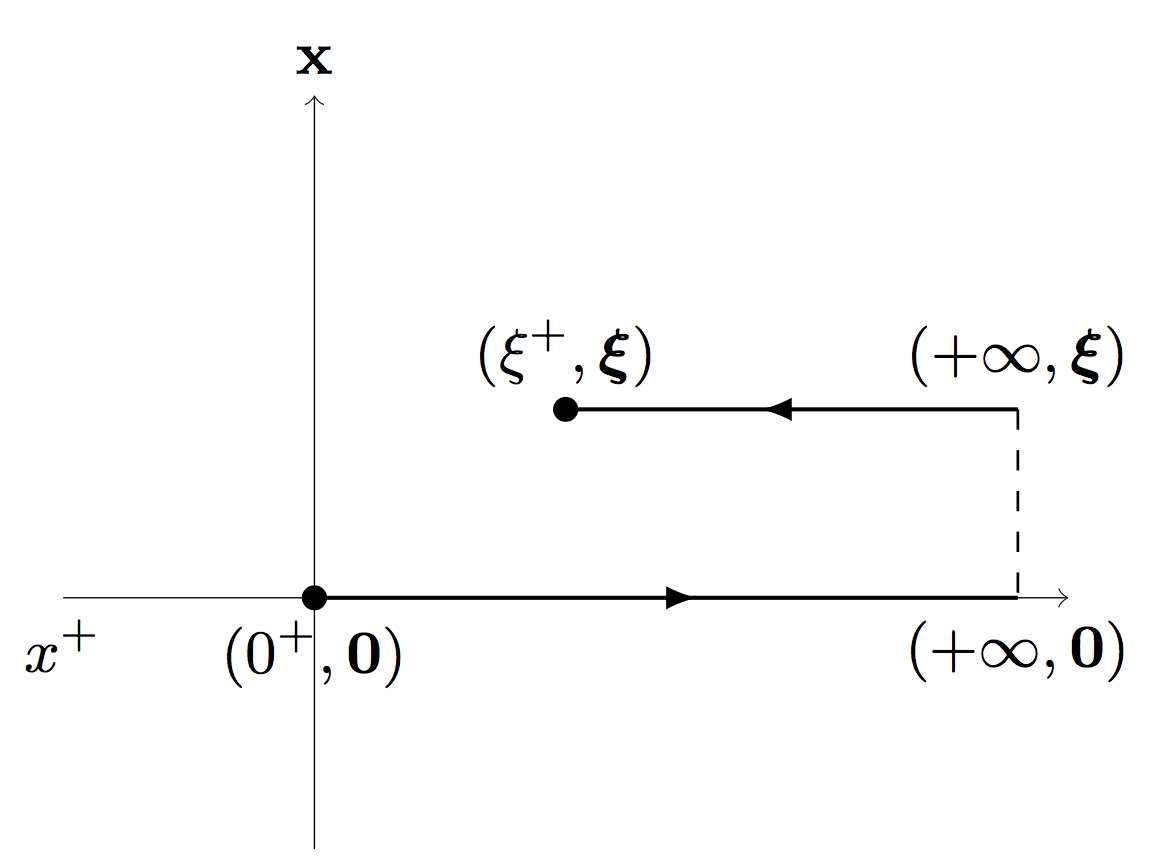}
 \end{center}
\caption{\label{fig:staple}\textquoteleft Staple' gauge link used in the gauge-invariant
definition of the Weizs\"acker-Williams gluon TMD.}

\end{figure}

In this short paragraph, we demonstrate the equivalence of the small-$x$ Weizs\"acker-Williams gluon distribution (for a left-moving hadron) as defined in Eq. (\ref{eq:F3}), and its standard operator definition \cite{Boer:2003cm,Dominguez:2011wm}:
\begin{equation}
\begin{aligned}\mathcal{F}_{gg}^{\left(3\right)}\left(x_2,k_{t}\right) & \equiv2\int\frac{\mathrm{d}^{3}\vec{\xi}}{\left(2\pi\right)^{3}p_{A}^{-}}e^{ix_2 p_{A}^{-}\xi^{+}}e^{-i\mathbf{k}_{t}\cdot\boldsymbol{\xi}}\mathrm{Tr}\Bigl\langle A\Bigr|F^{i-}(\vec{\xi})U^{\left[+\right]\dagger}_{\vec{\xi}}F^{i-}(\vec{0})U^{\left[+\right]}_{\vec{0}}\Bigl|A\Bigr\rangle\;,\end{aligned}
\label{eq:WeizsackerWilliamsdef}
\end{equation}
which is valid at all values of $x$, and where $U^{\left[+\right]}$ is a so-called staple gauge link (see Fig. \ref{fig:staple}):
\begin{equation}
\begin{aligned}U^{\left[+\right]}_{\vec{\xi}} & \equiv U\left(0^{+},+\infty;\boldsymbol{0}\right)U\left(+\infty,\xi^{+};\boldsymbol{\xi}\right)\;,\\
U^{\left[+\right]\dagger}_{\vec{\xi}} & \equiv U\left(\xi^{+},+\infty;\boldsymbol{\xi}\right)U\left(+\infty,0^{+};\boldsymbol{0}\right)\;,
\end{aligned}
\end{equation}
with
\begin{equation}
U\left(a,b;\mathbf{x}\right)\equiv\mathcal{P}e^{ig_{s}\int_{a}^{b}\mathrm{d}z^{+}A_{a}^{-}\left(z^{+},\mathbf{x}\right)t^{a}}\;.
\end{equation}

Setting $\mathrm{exp} \left(i x_2 p_A^- \xi^+ \right)\approx 1$ since $x_2$ is small, as well as making use of translational invariance and the fact that the light-cone Fock states are related
to the hadronic states $|A\rangle$ as follows (see \cite{Marquet:2016cgx,Dominguez:2011wm}):
\begin{equation}
\langle\mathcal{O}\rangle_{x_2}=\frac{\langle A|\mathcal{O}|A\rangle}{\langle A|A\rangle},\label{eq:Fock2hadronic}
\end{equation}
with the normalization $\langle A|A\rangle=\left(2\pi\right)^{3}2p_{A}^{-}\delta^{\left(3\right)}(\vec{0})$,
we obtain:
\begin{equation}
\begin{aligned}\mathcal{F}_{gg}^{\left(3\right)}\left(x_2,k_{t}\right) & =4\int\frac{\mathrm{d}^{3}\vec{v}\mathrm{d}^{3}\vec{w}}{\left(2\pi\right)^{3}}e^{-i\mathbf{k}_t\cdot\left(\mathbf{v}-\mathbf{w}\right)}\mathrm{Tr}\Bigl\langle F^{i-}\left(\vec{v}\right)U^{\left[+\right]\dagger}_{\vec{v}}F^{i-}\left(\vec{w}\right)U^{\left[+\right]}_{\vec{w}} \Bigr\rangle_{x_2}\;.\end{aligned}
\label{eq:WWFFFourier}
\end{equation}
The field tensor $F^{i-}=\partial_i A^-$ (in our choice of gauge) is related to the derivative of a Wilson line as follows:
\begin{equation}
\partial_{i}U_{\mathbf{x}}=ig_{s}\int\mathrm{d}x^{+}U\left(-\infty,x^{+};\mathbf{x}\right)F^{i-}\left(\vec{x}\right)U\left(x^{+},+\infty;\mathbf{x}\right)\;.
\end{equation}
Then, using the rules for the decomposition of Wilson lines, such as:
\begin{equation}
U(+\infty,w^+,\mathbf{w})=U(+\infty,-\infty,\mathbf{w})U(-\infty,w^+,\mathbf{w})\; ,
\end{equation}
the average in Eq. \eqref{eq:WWFFFourier} becomes
\begin{equation}
\aligned
&\int \mathrm{d}v^+ \mathrm{d}w^+\mathrm{Tr}\Bigl\langle F^{i-}\left(\vec{v}\right)U(v^+,+\infty,\mathbf{v}) U(+\infty,w^+,\mathbf{w})F^{i-}\left(\vec{w}\right) U(w^+,+\infty,\mathbf{w}) U(+\infty,v^+,\mathbf{v})\Bigr\rangle_{x_2}\;,\\
& =-\frac{1}{g_s^2} \mathrm{Tr}\Bigl\langle (\partial_{i} U_{\mathbf{v}})  U_{\mathbf{w}}^{\dagger} (\partial_{i} U_{\mathbf{w}}) U_{\mathbf{v}}^{\dagger})\Bigr\rangle_{x_2}\;,
\endaligned
\end{equation}
and one recovers expression (\ref{eq:F3}). The proof for the other gluon TMDs is analogous.

\bibliographystyle{apsrev4-1}
\bibliography{bibliography}

\end{document}